\documentclass[english,prd,superscriptaddress,nofootinbib,preprintnumbers,showpacs]{revtex4}

\usepackage[latin1]{inputenc}
\usepackage{graphicx}
\usepackage{amsmath,amsthm,amssymb}
\usepackage{bm}
\usepackage{amsfonts}
\usepackage{dcolumn}

\usepackage{color}

\newcommand{\beqa}{\begin{eqnarray}}
\newcommand{\eeqa}{\end{eqnarray}}
\newcommand{\simg}{\gtrsim}

\def\be{\begin{equation}}
\def\ee{\end{equation}}
\def\ba{\begin{eqnarray}}
\def\ea{\end{eqnarray}}
\def\Mpl{M_{\rm pl}}

\usepackage{color}

\usepackage{babel}
\begin{document}

\title{Implications of the B-mode Polarization Measurement for Direct Detection 
of Inflationary Gravitational Waves}

\author{Sachiko Kuroyanagi}
\affiliation{Department of Physics, Faculty of Science, 
Tokyo University of Science, 
1-3, Kagurazaka, Shinjuku-ku, Tokyo 162-8601, Japan}

\author{Shinji Tsujikawa}
\affiliation{Department of Physics, Faculty of Science, 
Tokyo University of Science, 
1-3, Kagurazaka, Shinjuku-ku, Tokyo 162-8601, Japan}

\author{Takeshi Chiba}
\affiliation{Department of Physics, College of Humanities and Sciences, 
Nihon University, Tokyo 156-8550, Japan}

\author{Naoshi Sugiyama}
\affiliation{Department of Physics, Nagoya
University, Chikusa, Nagoya 464-8602, Japan} 
\affiliation{Kavli Institute
for Physics and Mathematics of the Universe, University of Tokyo,
Chiba 277-8582, Japan}

\begin{abstract}

The prospects for direct measurements of inflationary 
gravitational waves by next generation interferometric detectors 
inferred from the possible detection of B-mode polarization of 
the cosmic microwave background are studied.  
We compute the spectra of the gravitational wave background and 
the signal-to-noise ratios by 
various types of interferometric detectors (DECIGO, BBO, and Ultimate DECIGO) 
for large-field inflationary models in which the tensor-to-scalar ratio 
is greater than the order of 0.01. 
If the reheating temperature $T_{\rm RH}$ of chaotic inflation with 
the quadratic potential is high ($T_{\rm RH}>7.9\times10^6$ GeV 
for upgraded DECIGO, 
$T_{\rm RH}> 1.8\times 10^{6}$ GeV for BBO,
and $T_{\rm RH}> 2.8\times 10^{3}$ GeV for Ultimate DECIGO), 
it will be possible to reach the sensitivity of the gravitational 
background in future experiments at $3\sigma$ confidence level. 
The direct detection is also possible for natural inflation with 
the potential $V(\phi)=\Lambda^4 [1-\cos(\phi/f)]$, provided 
that $f>4.2 M_{\rm pl}$ (upgraded DECIGO) 
and $f>3.6 M_{\rm pl}$ (BBO) with $T_{\rm RH}$ higher 
than $10^8$ GeV.
The quartic potential $V(\phi)=\lambda \phi^4/4$ 
with a non-minimal coupling $\xi$ between the inflaton 
field $\phi$ and the Ricci scalar $R$ gives 
rise to a detectable level of gravitational waves 
for $|\xi|$ smaller than the order of 0.01,
irrespective of the reheating temperature.

\end{abstract}

\date{\today}

\pacs{98.80.Cq, 95.30.Cq}

\maketitle

\section{Introduction}

The inflationary paradigm was originally proposed to address the horizon, flatness, 
and monopole problems in the standard big-bang cosmology \cite{inflation}.
Moreover, inflation can be responsible for the generation of primordial 
density perturbations \cite{scalar} and gravitational waves \cite{tensor}
by stretching quantum fluctuations over super-Hubble scales. 
The simplest slow-roll inflationary scenario driven by the potential energy 
of a single scalar field gives rise to nearly scale-invariant primordial 
power spectra of scalar and tensor perturbations. 

While the primordial scalar perturbation has been observed by
COBE \cite{cobe}, WMAP \cite{wmap}, and Planck \cite{planck} 
satellites as a form of temperature anisotropies of the Comic Microwave 
Background (CMB), gravitational waves had eluded the detection 
for a long time. This is attributed to the fact that the relative amplitude 
of gravitational waves to scalar perturbations is suppressed 
in the standard inflationary scenario (see Refs.~\cite{reviews} for reviews). 

The tensor-to-scalar ratio $r$ and the spectral index $n_s$ of 
scalar perturbations are two important quantities to 
probe models of inflation by using CMB observations.  
Although the Planck satellite measured $n_s$ in high precision, 
this is not sufficient to narrow down the allowed models of inflation unless $r$ is 
constrained tightly \cite{Planckinf,Martin,Tsuji13}. 
The detection of primordial gravitational waves is crucial to break 
the degeneracy between inflationary models.
In fact, there is the consistency relation between $r$ and 
the spectral index of primordial gravitational waves $n_t$: $r=-8n_t$ for potential-driven 
slow-roll inflation \cite{ll}. 
The detection of gravitational waves is a litmus test for inflationary cosmology.  

Recently, the BICEP2 collaboration reported the evidence for the detection 
of B-mode polarization in the CMB \cite{bicep2}. They claimed that 
the observed B-mode can be fit by a 
lensed $\Lambda$-Cold-Dark-Matter ($\Lambda$CDM) model with 
$r=0.2^{+0.07}_{-0.05}$. This bound is larger than the upper limit 
$r<0.11$ derived by the Planck data of temperature anisotropies \cite{planck,Planckinf}
combined with the WMAP large-angle polarization (WP) data \cite{WMAP9}. 
There is also an argument that the presence of the polarized dust 
weakens the constraints on $r$ \cite{Seljak}.
Since the BICEP2 measured the B-mode polarization at a single frequency (150 GHz), 
the upcoming independent observations at different frequencies 
with precise measurements of the dust polarization
will be crucial to clarify whether the signal 
is really the cosmological origin or not.

If the B-mode polarization reported by BICEP2 comes from the primordial origin, 
it marks a milestone in the inflationary cosmology \cite{Kami}. 
Since the primordial B-mode can be generated only from tensor or vector 
perturbations \cite{Polnarev} and vector perturbations sourced by topological defects 
are disfavored by the CMB data \cite{hindmarsh}, 
the detection of B-mode polarization indicates an indirect discovery of 
inflationary gravitational waves.   
A relatively large value of the tensor-to-scalar ratio of the order of 
0.1 can be explained by the so-called large field inflationary 
models in which the field variation  $\Delta \phi$ 
during inflation is greater than $10M_{\rm pl}$, where 
$M_{\rm pl}=2.435 \times 10^{18}$ GeV is 
the reduced Planck mass \cite{Lythbound}.

The typical examples of the large-field models are 
chaotic inflation \cite{chaotic} and natural inflation \cite{natural}. 
Although chaotic inflation with quartic potential is disfavored by  
the Planck data \cite{Planckinf},  
the non-minimal coupling \cite{FM,Fakir,Higgs} can reduce the value of $r$ \cite{nonper1,nonper2}
so that the model can be compatible with the data. 
A similar property also holds in the presence of Galileon
couplings \cite{Galileon} and field derivative couplings 
to the Einstein tensor \cite{derivative}.

If the large-field models correspond to a realistic paradigm of inflation, 
the next generation interferometric detectors such as DECIGO \cite{decigo} and BBO
\cite{bbo} may allow us to detect inflationary gravitational waves
directly.  
These instruments are designed to measure tensor
perturbations at much shorter wavelengths relative to the
perturbations associated with the CMB B-mode polarization.  These
frequencies correspond to the mode that re-entered the Hubble radius
during the radiation-dominated epoch.  The direct detection of
gravitational waves 
\cite{Turner:1996ck,Cooray:2005xr,Smith:2005mm,Smith:2006xf,Smith:2008pf,Chongchitnan:2006pe,Friedman:2006zt}
contains useful information of the very early
Universe, such as the effective relativistic degrees of freedom
\cite{watanabe,himemoto} and the reheating temperature 
after inflation \cite{yokoyama1,yokoyama2}.

In this paper, we compute the spectra of the 
gravitational wave background for several large-field inflationary models: chaotic inflation, 
natural inflation and non-minimally coupled inflation with quartic potential. 
Then, we update the previous calculations \cite{kuroyanagi} of 
signal-to-noise ratios by DECIGO and BBO with various reheating temperatures.
We also calculate signal-to-noise ratio for Ultimate DECIGO whose sensitivity 
is only limited by quantum noise \cite{decigo}. 
We show that, even if these models are degenerate in terms of 
the CMB observables $n_s$ and $r$, it is possible to distinguish 
them from the direct detection of inflationary gravitational waves.

This paper is organized as follows.
In Sec.~\ref{secback}, we present background equations of motion 
during inflation/reheating in the presence of non-minimal couplings.
In Sec.~\ref{secinf}, we review primordial gravitational waves 
generated during slow-roll inflation including non-minimal 
couplings between the inflaton field $\phi$ and the Ricci scalar $R$.
In Sec.~\ref{secspe}, we calculate the spectra of the 
gravitational background for the three large-field inflationary 
models by using the bounds derived from the recent CMB data. 
In Sec.~\ref{secsnr}, we study the detectability of inflationary 
gravitational waves by computing the signal-to-noise 
ratio (SNR) associated with upgraded DECIGO, 
BBO, and Ultimate DECIGO.
Section \ref{secsum} is devoted to summary.

\section{Background equations of motion during inflation and reheating}
\label{secback}

We start with the inflationary model given by the action 
\be
S=\int d^4x \sqrt{-g} \left[ \frac{M_{\rm pl}^2}{2}R
-\frac12 g^{\mu \nu}\partial_{\mu}\phi \partial_{\nu}\phi
-V(\phi)-\frac12 \xi \phi^2 R \right]\,,
\label{action}
\ee
where $R$ is the Ricci scalar of the metric $g_{\mu\nu}$, $V(\phi)$ is 
the potential of a scalar field $\phi$, and $\xi$ is 
the non-minimal coupling. 
In our sign convention, the conformal coupling 
corresponds to $\xi=1/6$.
We study the following three models in which 
the variation of the field during inflation is greater
than the order of $M_{\rm pl}$:
\ba
{\rm (i)} &&V=\frac{1}{2}m^2\phi^2\,,\qquad \xi=0\,, \label{mo1} \\
{\rm (ii)}&&V=\Lambda^4\left[1- \cos\left(\frac{\phi}{f}\right)\right]\,,
\qquad \xi=0\,,\\
{\rm (iii)}&&V=\frac14 \lambda \phi^4\,,\qquad 
\xi \neq 0\,,
\label{mo3}
\ea
where $m$, $\Lambda$, $f$, and $\lambda$ are constants.
The models (i) and (iii) belong to the class of chaotic inflation \cite{chaotic} 
with the quadratic potential and the quartic potential, respectively.
In the case (iii), we have introduced the non-minimal coupling $\xi$,
as this can reduce the tensor-to-scalar ratio $r$. 
The potential of the model (ii) corresponds to that of 
natural inflation \cite{natural}. 
The slow-roll inflation is possible for $f> M_{\rm pl}$.    
For large decay constant $f \gg M_{\rm pl}$, it is known that natural inflation 
is indistinguishable from quadratic chaotic inflation. 

\subsection{Background equations of motion}

We consider the flat Friedmann-Lema\^{i}tre-Robertson-Walker 
(FLRW) background described by 
the line-element $ds^2=-dt^2+a^2(t)\delta_{ij}dx^i dx^j$, 
where $a(t)$ is the scale factor with cosmic time $t$.
During the reheating after inflation, the inflaton energy density 
decays into the radiation energy density $\rho_r$.
The Born decay to light particles (decay constant $\Gamma$) 
during the oscillating stage of a scalar field can be effectively 
described by taking into account 
the friction term $\Gamma \dot{\phi}$ to the 
inflaton equation of motion \cite{Alb,Dolgov} (a dot 
represents a derivative with respect to $t$).
Then, the Friedmann equation and the scalar field equation 
of motion are given, respectively, by 
\ba
& &3H^2 (M_{\rm pl}^2-\xi \phi^2)=\frac12 \dot{\phi}^2
+V(\phi)+6H \xi \phi \dot{\phi}+\rho_r\,,\label{be1}\\
& &\ddot{\phi}+(3H+\Gamma) \dot{\phi}+V_{,\phi}
+6(2H^2+\dot{H}) \xi \phi=0\,,\label{be2}
\ea
where $V_{,\phi} \equiv dV/d\phi$ and $H \equiv \dot{a}/a$ 
is the Hubble parameter. We caution that Eq.~(\ref{be2}) 
with the $\Gamma \dot{\phi}$ term is valid only during 
the oscillating stage of the inflaton around the potential 
minimum \cite{Kofman}.
Provided that $H \gg \Gamma$ during inflation, the decay 
term $\Gamma \dot{\phi}$ does not play any significant 
role until the onset of reheating.

Due to the energy conservation, the radiation density $\rho_r$ 
obeys the equation of motion 
\be
\dot{\rho}_r+4H\rho_r=\Gamma \dot{\phi}^2\,.
\label{rhoeq}
\ee
From Eq.~(\ref{be1}), we obtain
\be
H=\frac{6 \xi \phi \dot{\phi}+
\sqrt{6[2(V+\rho_r)(M_{\rm pl}^2-\xi \phi^2)+\dot{\phi}^2
\{ M_{\rm pl}^2+\xi \phi^2 (6\xi-1)\} ]}}
{6(M_{\rm pl}^2-\xi \phi^2)}\,.
\label{Hu}
\ee
Taking the time derivative of Eq.~(\ref{be1}) and 
eliminating the term $\dot{H}$ on account of 
Eq.~(\ref{be2}), we have 
\ba
\ddot{\phi}
&=&
-\{M_{\rm pl}^2 [ (3H+\Gamma)\dot{\phi} +V_{,\phi}]
-\phi[ 3\dot{\phi}^2-12H^2M_{\rm pl}^2
+(3H+\Gamma)\phi \dot{\phi}+\phi V_{,\phi}
+4\rho_r ]\xi \nonumber \\
& &
+6\phi (\dot{\phi}-2H \phi) (\dot{\phi}+H \phi)\xi^2\}
/[M_{\rm pl}^2+\xi \phi^2 (6\xi-1)]\,.
\label{ddotphi}
\ea
We solve the differential equations (\ref{rhoeq}) and 
(\ref{ddotphi}) with Eq.~(\ref{Hu}).

For the models (i) and (ii), the slow-roll inflationary stage is 
followed by oscillations of the massive inflaton field. 
This corresponds to the temporal matter era
during which the evolution of the scale factor is given 
by $a \propto t^{2/3}$. 
Around the time $t_{\rm RH} \simeq \Gamma^{-1}$, 
$\rho_r$ catches up with the inflaton energy density 
$\rho_{\phi}=\dot{\phi}^2/2+V(\phi)$ \cite{Kolb}.
Here, the subscript ``RH" denotes the value at the end of reheating.
We numerically find that the radiation energy density at $t=t_{\rm RH}$ is 
given by $\rho_r (t_{\rm RH}) \simeq 0.5\,\Gamma^2 M_{\rm pl}^2$. 
On using the relation $\rho_r (T)=(\pi^2/30)g_{*}(T)T^4$, where 
$T$ is the temperature and
$g_*$ is the number of relativistic degree of freedom, 
the reheating temperature is given by
\be
T_{\rm RH} \simeq 1.1\,g_{*,{\rm RH}}^{-1/4}\, 
(\Gamma M_{\rm pl})^{1/2}\,.
\label{reheat}
\ee
We take $g_{*,{\rm RH}}=106.75$ for the value 
of $g_*$ at the end of reheating.
For $t>t_{\rm RH}$, the Universe enters the 
radiation-dominated epoch.

\subsection{Dynamics of reheating for the model (iii) with $|\xi| \ll 1$}

In the model (iii), the CMB observables $n_s$ and $r$ are subject to 
change relative to the case $\xi=0$ because the effect of non-minimal 
couplings cannot be neglected due to the large inflaton value.
As we will see in Sec.~\ref{BICEPcon}, the model is compatible 
with the observational data even for $|\xi| \ll 1$.
Since the amplitude of the field $\phi$ drops below $M_{\rm pl}$ 
after inflation, the effect of non-minimal couplings on the background 
Eqs.~(\ref{rhoeq})-(\ref{ddotphi}) should be negligible during 
reheating for $|\xi| \ll 1$. 
In this case, the dynamics of reheating is driven by a massless 
inflaton field, so that the scale factor evolves as $a \propto t^{1/2}$ 
and hence, there is no transient matter era between inflation 
and the radiation-dominated epoch.

For the model (iii) with $|\xi| \ll 1$, the dynamics of reheating can 
be analytically known by using the virial theorem 
$\langle \dot{\phi}^2/2 \rangle=2 \langle V(\phi) \rangle$, 
where $\langle \cdots \rangle$ corresponds to the time 
average over inflaton oscillations.
Dropping the contribution of non-minimal couplings 
in Eq.~(\ref{be2}), it follows that 
\be
\langle \dot{\rho}_{\phi} \rangle +\left(4H+4\Gamma/3 \right)
\langle \rho_{\phi} \rangle \simeq 0\,.
\label{rhophi}
\ee
Integration of Eq.~(\ref{rhophi}) gives 
\be
\langle \rho_{\phi} \rangle=
\rho_{\phi i} \left( \frac{t_i}{t} \right)^2
e^{-(4/3)\Gamma (t-t_i)}\,,
\label{rhophi2}
\ee
where the subscript ``$i$'' is used for quantities 
at the onset of inflaton oscillations and we used 
the solution $a=a_i (t/t_i)^{1/2}$ for $t>t_i$.
Substituting Eq.~(\ref{rhophi2}) into Eq.~(\ref{rhoeq}), i.e.,
$\langle \dot{\rho}_{r} \rangle
+4H \langle \rho_{r} \rangle=(4/3)\Gamma 
\langle \rho_{\phi} \rangle$, we obtain the 
following solution
\be
\langle \rho_{r} \rangle=
\rho_{\phi i} \left( \frac{t_i}{t} \right)^2
\left[ 1-e^{-(4/3)\Gamma (t-t_i)} \right]\,.
\label{rhor2}
\ee

The time $t_i$ can be estimated as $t_i \simeq 1/(2H_i)$. 
Provided that $\Gamma \ll H_i$,
the time $t_{\rm RH}$ at which $\langle \rho_{r} \rangle$ 
equals $\langle \rho_{\phi} \rangle$ is
given by $t_{\rm RH} \simeq 0.52 \Gamma^{-1}$. 
On using the relation $\rho_{\phi i}  \simeq 3M_{\rm pl}^2H_i^2$, 
we have $\langle \rho_{r} \rangle (t_{\rm RH}) \simeq 
1.4\,\Gamma^2 M_{\rm pl}^2$.
Equating this with the radiation density
$(\pi^2/30)g_{*,{\rm RH}}T_{\rm RH}^4$, it follows that 
\be
T_{\rm RH} \simeq 1.4\,g_{*,{\rm RH}}^{-1/4}\, 
(\Gamma M_{\rm pl})^{1/2}\,.
\label{reheat2}
\ee
We have numerically solved (\ref{rhoeq}) and 
(\ref{ddotphi}) with Eq.~(\ref{Hu}) from the onset of reheating 
and confirmed that the above analytic estimation shows 
good agreement with our numerical results for $|\xi| \ll 1$.

In Sec.~\ref{secsnr}, we employ the relations (\ref{reheat}) and (\ref{reheat2})
to estimate $T_{\rm RH}$ for a given decay constant $\Gamma$.
As we see in Sec.~\ref{secspe}, the absence of the temporal matter era
for the model (iii) affects the resulting spectrum 
of the gravitational wave background relative to the models (i) and (ii).

\subsection{The number of e-foldings during inflation}

The number of e-foldings during inflation is defined by 
$N(k) \equiv \ln (a_{\rm end}/a_k)$, where $a_k$ is 
the value of $a$ when a wave number $k$ equals $aH$ 
during inflation and $a_{\rm end}$ is its value at the end 
of inflation. This quantity is related to the observables such as 
$n_s$ and $r$, so the accurate estimation of $N(k)$ 
is necessary to place concrete constraints on inflationary models.
After inflation, there is a reheating stage followed by the 
radiation-dominated epoch. 
Then, the wave number $k=a_kH_k$ divided by $a_0H_0$ 
(the label ``0'' represents today's values) reads 
\be
\frac{k}{a_0H_0}=e^{-N(k)} 
\frac{a_{\rm end}}{a_{\rm RH}}
\frac{a_{\rm RH}}{a_{0}} \frac{H_k}{H_{0}}\,,
\label{kaH}
\ee
where we express the Hubble constant as 
$H_0=2.133h \times 10^{-42}$ GeV and $a_0H_0=2.235\times 10^{-4}(h/0.67){\rm Mpc}^{-1}$. 

We assume that the entropy at the end of reheating 
(with the relativistic degrees of freedom 
$g_{s,{\rm RH}}$) is conserved in the photon and neutrino background 
today. This leads to the following relation \cite{Mark}:
\be
g_{s,{\rm RH}}T_{\rm RH}^3 a_{\rm RH}^3=
\left[ 2T_0^3+\frac78 \cdot 6\,T_{\nu 0}^3 \right]a_0^3\,,
\ee
where $T_{0}=2.725\,{\rm K} = 2.348 \times 10^{-13}$~GeV 
is the CMB temperature at present and today's temperature 
of neutrinos is given by $T_{\nu 0}=(4/11)^{1/3}T_0$. 
Then, we obtain 
\be
\frac{a_0}{a_{\rm RH}}=\left( \frac{11}{43} g_{s,{\rm RH}}
\right)^{1/3} \frac{T_{\rm RH}}{T_0}\,.
\ee
If the total energy density $\rho$ during reheating is given by 
$\rho \propto a^{-q}$, where $q$ is constant, 
the number of e-foldings in the reheating 
period can be estimated as 
$N_{\rm RH} \equiv \ln (a_{\rm RH}/a_{\rm end})
=(1/q) \ln (\rho_{\rm end}/\rho_{\rm RH})$.
The energy density $\rho_{\rm RH}$ is related to 
the reheating temperature $T_{\rm RH}$, as
$\rho_{\rm RH}=(\pi^2/30)g_{*,{\rm RH}}T_{\rm RH}^4$, 
where $g_{*,{\rm RH}}$ is the number of relativistic degrees 
of freedom at the end of reheating.

On using the aforementioned relations in Eq.~(\ref{kaH}),
it follows that 
\be
N(k)=-\ln \left( \frac{k}{a_0H_0} \right)-\frac{1}{q} \ln \left( 
\frac{30}{\pi^2} \right)-\frac13 \ln \left( \frac{11}{43} \right)
+\ln \left( \frac{g_{*,{\rm RH}}^{1/q}}{g_{s,{\rm RH}}^{1/3}}
\right)
-\ln \left( \frac{\rho_{\rm end}^{1/q}}{T_{\rm RH}^{4/q-1}T_0} 
\right)+\ln \left( \frac{H_k}{H_0} \right)\,.
\label{Nesti2}
\ee
Here the Hubble parameter $H_k$ is associated with the inflationary energy 
density $\rho_{\rm inf}$, as $H_k=\sqrt{\rho_{\rm inf}/(3M_{\rm pl}^2)}$.
Since $\rho_{\rm end}$ differs from $\rho_{\rm inf}$, we define 
\be
\alpha \equiv \frac{\rho_{\rm end}}{\rho_{\rm inf}}\,,
\ee
to quantify their difference. 
In the following, we set  $g_{*,{\rm RH}}=g_{s,{\rm RH}}$ by assuming 
that no entropy production occurs after the reheating stage 
until the neutrino decoupling ($\sim$ MeV).   
Depending on the values of $q$,
we shall discuss two qualitatively different cases.

\subsubsection{Massive inflaton $(q=3)$}

The models (i) and (ii) belong to the class in which the 
energy density during reheating decreases as $\rho \propto a^{-3}$. 
In this case, the number of e-foldings (\ref{Nesti2}) reads
\be
N(k) = 55.9-\frac13 \ln \alpha-\ln \left( \frac{k}{a_0H_0} \right) 
-\ln \left( \frac{h}{0.67} \right)+
\frac13 \ln \left( \frac{T_{\rm RH}}{10^9~{\rm GeV}} \right)
+\frac23 \ln \left( \frac{\rho_{\rm inf}^{1/4}}
{10^{16}~{\rm GeV}} \right)\,.
\label{efoldq=3}
\ee
As we will see in Sec.~\ref{secinf}, the Hubble parameter $H_k$ is 
related to the amplitude ${\cal P}_T$ of tensor perturbations as 
${\cal P}_T=2H_k^2/(\pi^2 M_{\rm pl}^2)=r{\cal P}_{\cal R}$, 
where ${\cal P}_{\cal R} \simeq 2.198 \times 10^{-9}$ 
is the amplitude of curvature perturbations constrained 
by Planck \cite{planck}. 
Then, the inflationary energy scale can be generally estimated as
\be
\rho_{\rm inf}^{1/4}=1.84\times  10^{16} 
\left(\frac{r}{0.1}\right)^{1/4} ~{\rm GeV} \,.
\label{rhoinf}
\ee

\begin{figure}
\includegraphics[height=3.3in,width=4.8in]{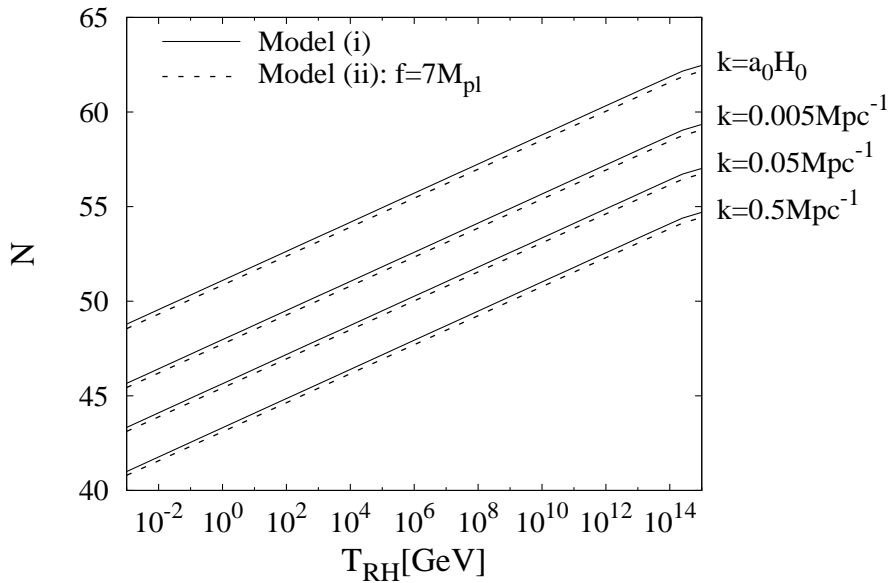}
\caption{\label{fig1}
The number of e-foldings $N(k)$ versus the reheating temperature 
$T_{\rm RH}$ for the model (i) (solid line) and the model (ii) 
with $f=7M_{\rm pl}$ (dashed line). We choose four different values 
of $k$ which are inside the observed CMB range. 
For increasing $T_{\rm RH}$ and decreasing $k$, $N(k)$ gets larger.}
\end{figure}

For the monomial potential $V(\phi)=\lambda \phi^n/n$ the field 
value during inflation is analytically known as
$\phi\simeq \sqrt{2n(N+n/4)}M_{\rm pl}$, with 
$\phi_{\rm end}=nM_{\rm pl}/\sqrt{2}$. 
Hence the parameter $\alpha$ can be estimated as
\be
\alpha \simeq \frac43 \frac{\phi_{\rm end}^n}{\phi^n} 
\simeq \frac43 \left( \frac{n}{4N} \right)^{n/2}\,,
\label{alpha}
\ee
where the factor $4/3$ comes from the contribution of the 
inflaton kinetic energy to $\rho_{\rm end}$. 
For the model (i), i.e., $n=2$, we have $\alpha \simeq 2/(3N)$. 
This leads to a non-negligible change to $N(k)$ of the order of 1.
Accordingly, Eq.~(\ref{efoldq=3}) becomes 
\be
N(k)=56.5+\frac13 \ln N(k)-\ln \left( \frac{k}{a_0H_0} \right)
-\ln \left( \frac{h}{0.67} \right)
+\frac13 \ln \left( \frac{T_{\rm RH}}{10^9~{\rm GeV}} \right)
+\frac16 \ln\left(\frac{r}{0.1}\right)\,.
\label{efoldq=3d}
\ee
For given $k$, $h$, and $T_{\rm RH}$, $N(k)$ is known by solving 
Eq.~(\ref{efoldq=3d}).
In Fig.~\ref{fig1}, we plot $N(k)$ versus $T_{\rm RH}$ for 
several different values of $k$.
This result is obtained by numerically solving the background 
equations of motion from inflation to the present epoch. 
We have confirmed that the numerical values of $N(k)$ show 
good agreement with those estimated by Eq.~(\ref{efoldq=3d}). 
On the largest scale observed in the CMB ($k=a_0H_0$), 
the number of e-foldings is in the range 
$53<N(k)<61$ for $10^3~{\rm GeV}<T_{\rm RH}<10^{13}~{\rm GeV}$.
For larger $k$, $N(k)$ becomes smaller.

In natural inflation (the model (ii)), the number of e-foldings 
can also be calculated as
\be
N \simeq -2\left(\frac{f}{M_{\rm pl}}\right)^2
\ln \left[ \sqrt{1+\frac12\left(\frac{M_{\rm pl}}{f}\right)^2}
\cos \left( \frac{\phi}{2f} \right)\right]\,,
\ee
with $\cos^2(\phi_{\rm end}/2f)=1/[1+(M_{\rm pl}/f)^2/2] $. 
Then, the parameter $\alpha$ reads
\be
\alpha \simeq \frac43\frac{\sin^2(\phi_{\rm end}/2f) }{\sin^2(\phi/2f)}\simeq 
\frac{4/3}{1+2(f/M_{\rm pl})^2\left(1-e^{-NM_{\rm pl}^2/f^2}\right)}.
\ee
Accordingly, Eq.~(\ref{efoldq=3}) is replaced by 
\ba
N(k)
&=& 56.5+\frac13 \ln \left[ \frac12+
\left( \frac{f}{M_{\rm pl}} \right)^2
\left(1-e^{-N(k)M_{\rm pl}^2/f^2}\right)\right]
-\ln \left( \frac{k}{a_0H_0} \right)
-\ln \left( \frac{h}{0.67} \right) \nonumber \\
& &+\frac13 \ln \left( \frac{T_{\rm RH}}{10^9~{\rm GeV}} \right)
+\frac16 \ln\left(\frac{r}{0.10}\right)\,.
\label{efoldq=3dnatural}
\ea
In the $f \to \infty$ limit, natural inflation becomes 
indistinguishable from quadratic chaotic inflation, so that 
Eq.~(\ref{efoldq=3dnatural}) reduces to Eq.~(\ref{efoldq=3d}). 

In Fig.~\ref{fig1}, we show numerically derived values of $N(k)$ 
versus $T_{\rm RH}$ for the model (ii) with $f=7M_{\rm pl}$. 
The number of e-foldings exhibits only a tiny difference from 
that in the model (i).

\subsubsection{Massless inflaton $(q=4)$}

For $|\xi| \ll 1$ the model (iii) belongs to the class of massless inflaton 
characterized by $q=4$. In this case, the number of e-foldings (\ref{Nesti2}) 
does not depend on $T_{\rm RH}$. 
Following the same argument as before and using the relations 
(\ref{rhoinf}) and (\ref{alpha}) with $n=4$, we obtain 
\be
N(k) = 61.6+\frac12 \ln N(k) -\ln \left( \frac{k}{a_0H_0} \right) 
-\ln \left( \frac{h}{0.67} \right)
-\frac{1}{12} \ln \left( \frac{g_{*,{\rm RH}}}{106.75} \right)
+\frac14 \ln \left( \frac{r}{0.1} \right)\,.
\label{efoldq=4}
\ee
If $h=0.67$, $g_{*,{\rm RH}}=106.75$ and $r=0.1$, 
we have $N(k)=63.6$ for $k=a_0H_0$ and 
$N(k)=58.2$ for $k=0.05$ Mpc$^{-1}$.

\section{Primordial perturbations generated in large-field inflationary models 
and constraints from CMB}
\label{secinf}

We consider the following perturbed metric on the 
flat FLRW background \cite{Bardeen}:
\be
ds^2=-(1+2A)dt^2+2\partial_i B dt dx^i+a^2(t)
\left[ (1+2\psi) \delta_{ij}+h_{ij} \right] dx^i dx^j\,,
\label{permet}
\ee
where $A$, $B$, and $\psi$ are scalar perturbations, and 
$h_{ij}$ is the tensor perturbation.
We choose the unitary gauge in which the field perturbation 
$\delta \phi$ vanishes, such that the gauge-invariant 
curvature perturbation ${\cal R} \equiv \psi
-H \delta \phi/\dot{\phi}$ is identical to $\psi$.
This fixes the time component of the gauge-transformation 
vector $\xi^{\mu}$. We have also fixed the spatial component 
of $\xi^{\mu}$ by gauging away the scalar perturbation $E$ appearing 
as the form $E_{,ij}$ in the metric (\ref{permet}).

\subsection{Tensor perturbations}

Expanding the action (\ref{action}) for the transverse and traceless 
tensor perturbation, the second-order action reads
\be
S_t^{(2)}=\int d^4 x\,a^3 \frac{M_{\rm pl}^2F}{4}
\left[ \frac12 \dot{h}_{ij}^2-\frac{1}{2a^2} 
(\partial h_{ij})^2 \right]\,,
\label{secondten}
\ee
where $F \equiv 1- \xi \phi^2/M_{\rm pl}^2$.
We write $h_{ij}$ in terms of Fourier components, as
\be
h_{ij}(t,{\bm x})=\sum_{\mu=+,\times}^{}
\int\frac{d^3k}{(2\pi)^{3/2}}\epsilon_{ij}^{\mu}
({\bm k})h_{\bm k}^{\mu}(t)e^{i{\bm k}\cdot {\bm x}}\,,
\label{hij}
\ee
where ${\bm k}$ is a comoving wave number.
The polarization tensors $\epsilon_{ij}^{+,\times}$ satisfy
symmetric and transverse-traceless conditions and are normalized as
$\sum_{i,j}^{}\epsilon_{ij}^{\mu}(\epsilon_{ij}^{\mu^{\prime}})^*
=2\delta^{\mu \mu^{\prime}}$. 
From the action (\ref{secondten}), the Fourier mode 
$h_{\bm k}^{\mu}$ obeys the equation of motion 
\be
\ddot{h}_{\bm k}^{\mu}+\left( 3H+\frac{\dot{F}}{F} \right)
\dot{h}_{\bm k}^{\mu}
+\frac{k^2}{a^2}h_{\bm k}^{\mu}=0\,.
\label{heq2}
\ee

For the quantization procedure, we express $h_{\bm k}^{\mu}$ 
in the form
\be
h_{\bm k}^{\mu}(t)=h_{k} (t)a_{\mu} ({\bm k})+
h_{k}^* (t)a_{\mu}^\dagger (-{\bm k})\,,
\ee
where the annihilation and creation operators 
satisfy $[a_{\mu} ({\bm k}), a_{\mu'}^{\dagger} ({\bm k}')]
=\delta_{\mu \mu'} \delta^{(3)}({\bm k}-{\bm k}')$.
A canonically normalized field, which is defined by
\be
v_{k} \equiv z h_{k}\,,\qquad {\rm where} \qquad
z \equiv aM_{\rm pl} \sqrt{F/2}\,, 
\ee
obeys the equation of motion 
\be
v_{k}''+\left( k^2-\frac{z''}{z} \right) v_k=0\,,
\label{vkeq}
\ee
where a prime represents a derivative with respect to 
the conformal time $\tau \equiv \int a^{-1}dt$. 
In the asymptotic past, the solution 
corresponding to the Bunch-Davies vacuum is given by 
$v_k= e^{-i k\tau}/\sqrt{2k}$, i.e., 
\be
h_k =\frac{e^{-ik \tau}}
{aM_{\rm pl}\sqrt{Fk}}\,,\qquad {\rm for} \qquad
\tau \to -\infty\,.
\label{initial}
\ee
On the quasi de Sitter background with a nearly constant Hubble parameter, 
the variation of the quantity $F$ is negligibly small relative to that of $a$, 
such that $z''/z \simeq a''/a \simeq 2/\tau^2$. 
Under this approximation, the solution to Eq.~(\ref{vkeq}) 
during inflation reads
\be
h_k (\tau) \simeq \frac{iH e^{-i k \tau}}{k^{3/2}M_{\rm pl} \sqrt{F}}
(1+ik \tau)\,.
\label{hktau}
\ee

We define the tensor power spectrum, as
\be
{\cal P}_T \equiv \frac{k^3}{\pi^2} \sum_{\mu=+,\times} 
\left| h_{\bm k}^{\mu} \right|^2\,.
\label{PTdef}
\ee
After the Hubble radius crossing ($k=aH$), the solution (\ref{hktau}) 
approaches $h_k \to iH/(k^{3/2}M_{\rm pl} \sqrt{F})$ so that
we obtain the primordial power spectrum 
\be
{\cal P}_T^{\rm prim} \simeq \frac{2H^2}
{\pi^2 M_{\rm pl}^2F}\biggr|_{k=aH}\,.
\label{PT2}
\ee
The formula (\ref{PT2}) has been derived under the condition 
that the variations of $H$ and $F$ are negligible during inflation. 
This is a good approximation for the perturbations relevant to 
CMB anisotropies ($50\sim 60$ e-foldings before the end of 
inflation), but it is not so for the modes associated with 
the direct detection of gravitational waves 
($10\sim 20$ e-foldings before the end of inflation). 
For precise computations of the spectrum of the gravitational 
wave background, we shall numerically solve Eq.~(\ref{heq2}) under 
the initial condition (\ref{initial}) without using the formula (\ref{PT2}).

\subsection{Scalar perturbations and the tensor-to-scalar ratio}

For the study of scalar perturbations and the tensor-to-scalar ratio, 
it is convenient to perform the so-called conformal transformation 
$\hat{g}_{\mu \nu}=\Omega g_{\mu \nu}$, where a hat 
is used for quantities in the new frame and $\Omega$ is 
a conformal factor. For the choice $\Omega=F(\phi)$, 
the action (\ref{action}) transforms to \cite{Maeda}
\be
\hat{S}=\int d^{4}x\sqrt{-\hat{g}}\left[\frac{1}{2}\Mpl^{2}\hat{R}
-\frac{1}{2}\hat{g}^{\mu\nu}\partial_{\mu}\chi\partial_{\nu}\chi
-U(\chi)\right]\,,
\label{Eaction}
\ee
where 
\be
U=\frac{V}{F^{2}}\,,\qquad\chi \equiv \int {\cal B}(\phi)\, d\phi\,,
\qquad {\cal B}(\phi) \equiv 
\sqrt{\frac{3}{2}\left(\frac{\Mpl F_{,\phi}}{F}\right)^{2}
+\frac{1}{F}}\,.
\label{Ure}
\ee
The perturbed metric in the Einstein frame (\ref{Eaction}) is 
given by $d\hat{s}^2=F ds^2$, where $ds^2$ corresponds 
to the metric (\ref{permet}). Decomposing the quantity $F$ into 
the background and perturbed components as 
$F(t)+\delta F(t,{\bm x})$, we obtain 
the following correspondence: 
\ba
& & \hat{a}=a\sqrt{F}\,,\qquad \hat{t}=\int \sqrt{F}dt\,,\qquad
\hat{H}=\frac{1}{\sqrt{F}} \left( H+\frac{\dot{F}}{2F} \right)\,,\label{conre1}\\
& & \hat{A}=A+\frac{\delta F}{2F}\,,\qquad \hat{B}=B\,,\qquad
\hat{\psi}=\psi+\frac{\delta F}{2F}\,,\qquad
\hat{h}_{ij}=h_{ij}\,,\label{conre2}
\qquad 
\ea
so that the tensor perturbation is invariant under the 
conformal transformation. 
On using the relations (\ref{conre1}), the power spectrum (\ref{PT2}) reads 
\be
{\cal P}_T^{\rm prim} \simeq \frac{2\hat{H}^2}
{\pi^2 M_{\rm pl}^2}\biggr|_{k=\hat{a}\hat{H}}\,,
\ee
at leading order in slow-roll (under which the term $\dot{F}/(2F)$ 
is negligible relative to $H$).

Using the correspondence (\ref{conre2}), one can show that 
the curvature perturbation ${\cal R}=\psi-H \delta \phi/\dot{\phi}$ 
is invariant under the conformal transformation \cite{nonper1,nonper2}. 
In the Einstein frame the power spectrum of ${\cal R}$ is equivalent 
to the one in standard slow-roll inflation, so it is given by \cite{reviews}
\be
{\cal P}_{\cal R}^{\rm prim} \simeq \frac{\hat{H}^2}
{8\pi^2 \epsilon_U M_{\rm pl}^2} \biggr|_{k=\hat{a}\hat{H}}\,,
\ee
at leading order in slow-roll, 
where $\epsilon_U \equiv (M_{\rm pl}^2/2)(U_{,\chi}/U)^2$. 
The scalar spectral index $n_s \equiv 1
+d\ln {\cal P}_{\cal R}^{\rm prim}/d\ln k|_{k=\hat{a}\hat{H}}$
and the tensor-to-scalar ratio 
$r={\cal P}_{T}^{\rm prim}/{\cal P}_{\cal R}^{\rm prim}|_{k=\hat{a}\hat{H}}$ 
are
\be
n_s=1-6\epsilon_U+2\eta_U\,,\qquad r=-8n_t
=16\epsilon_U\,,
\label{nsr}
\ee
where $\eta_U \equiv M_{\rm pl}^2 U_{,\chi \chi}/U$ and 
$n_t \equiv d\ln {\cal P}_{T}^{\rm prim}/d\ln k|_{k=\hat{a}\hat{H}}$ 
is the tensor spectral index. 
For a given inflaton potential, these observables can be 
explicitly expressed as a function of $\phi$ by 
using the relations (\ref{Ure}).

\subsection{Observational implications from the recent CMB data}
\label{BICEPcon}

We compute the CMB observables (\ref{nsr}) for the models (i)-(iii) 
to confront them with observations. 
We refer the readers to Refs.~\cite{Tsuji13} for detailed calculations 
of $n_s$ and $r$.
The number of e-foldings during inflation can be expressed 
as the integrated form $N(t)=-\int^{t}_{t_f} H (\tilde{t})d \tilde{t}$ 
in the Jordan frame. 
With a proper choice of a reference length scale, the number of 
e-foldings is a frame-independent quantity \cite{Catena}. 
From the third relation of Eq.~(\ref{conre1}), it follows that 
\be
N=\int_{\chi_f}^{\chi} \frac{U}{M_{\rm pl}^2 U_{,\chi}} d\chi
+\frac12 \ln \frac{F(t)}{F(t_f)}\,,
\label{efold}
\ee
where we have employed the slow-roll approximation in 
the Einstein frame.

\begin{figure}
\includegraphics[height=3.5in,width=3.5in]{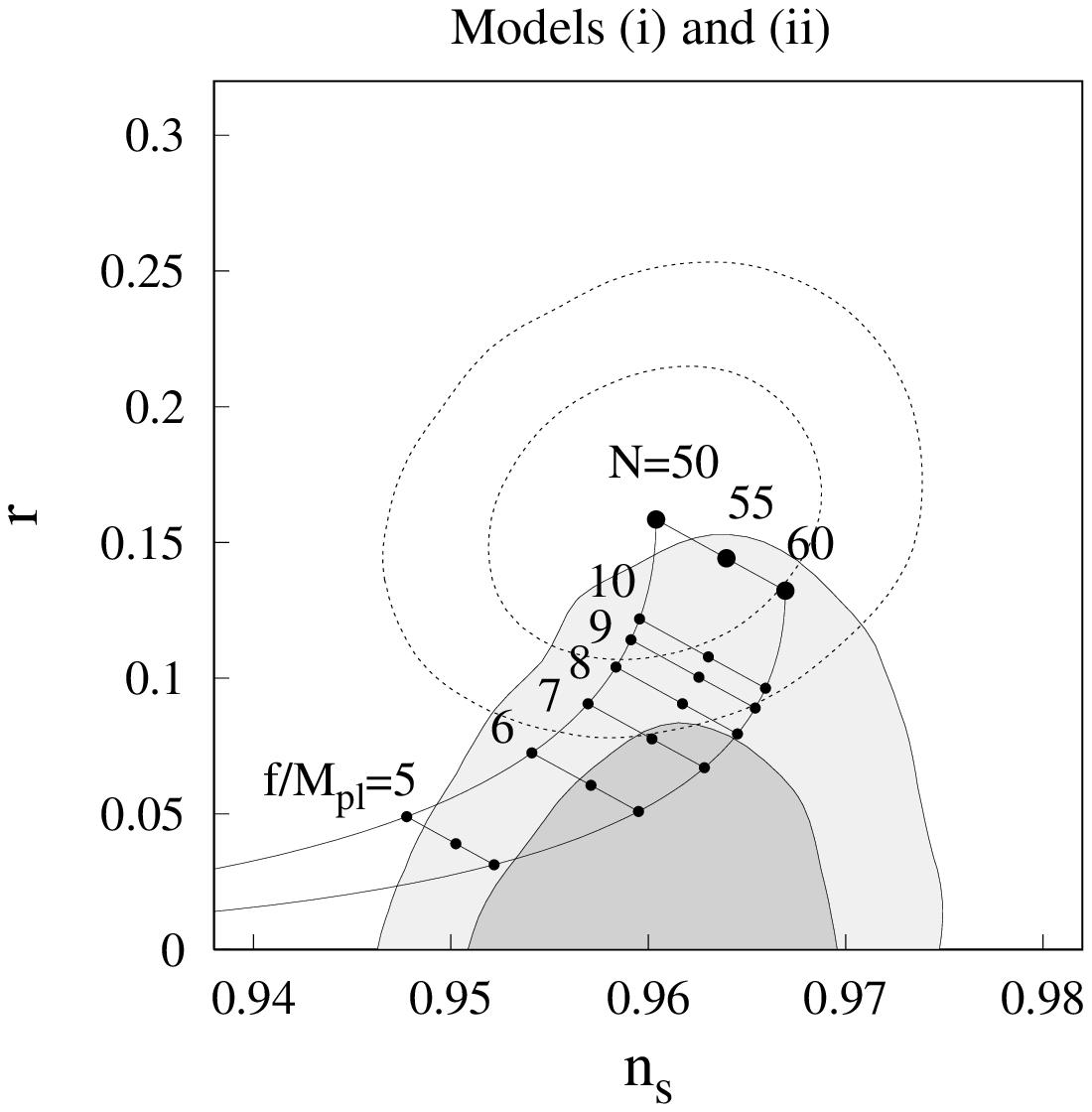}
\includegraphics[height=3.5in,width=3.5in]{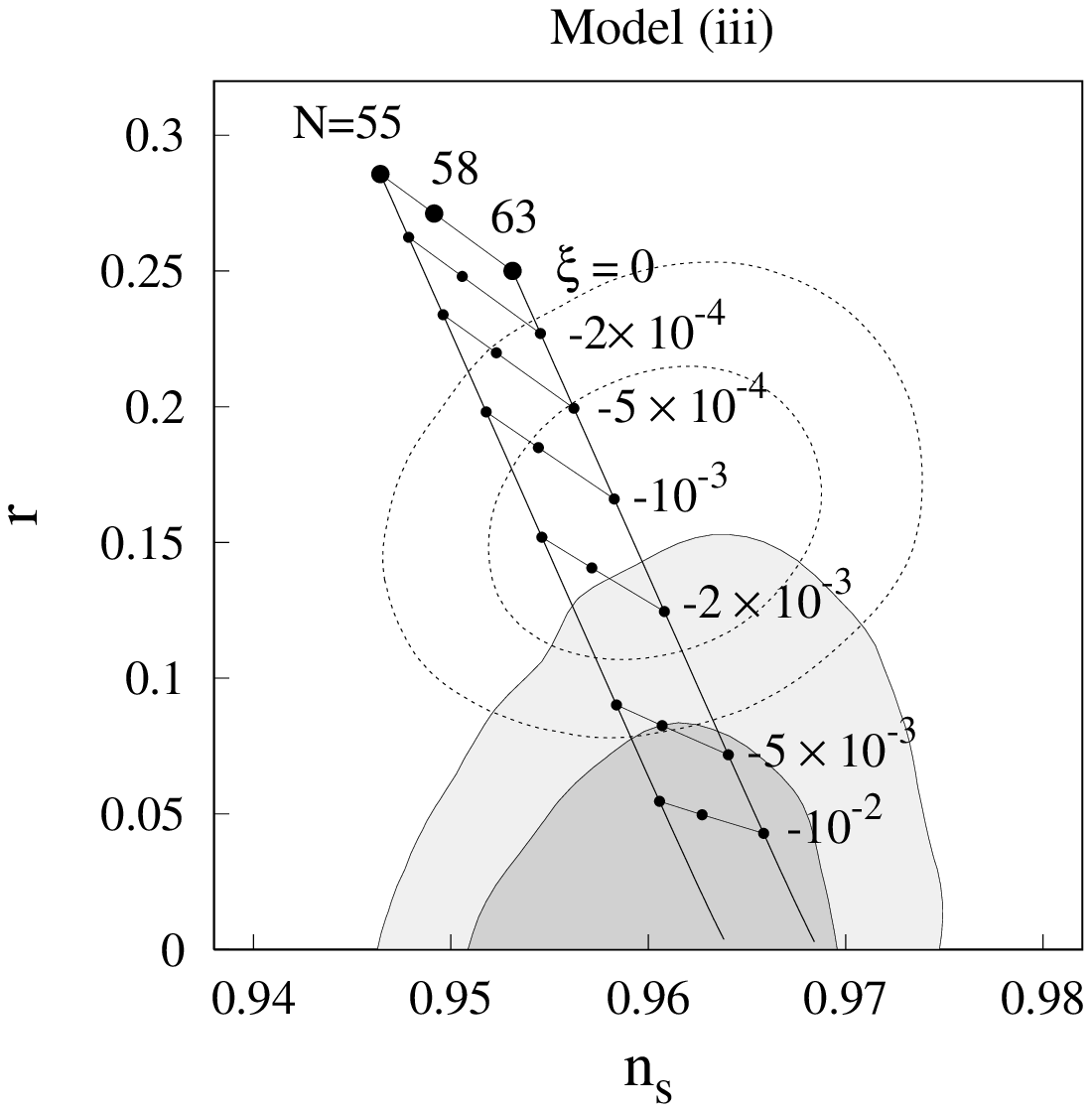}
\caption{\label{fig2} 
Two-dimensional observational constraints on the
models (i) and (ii) (left), and the model (iii) (right) in the
$(n_s,r)$ plane with the pivot wave number $k_*=0.05$ Mpc$^{-1}$.
The gray shaded regions represent the 68 \% C.L. (inside) and 
95 \% C.L. (outside) parameter spaces constrained by the joint data
analysis of Planck+WP+BAO+high-$\ell$.  We also show the 68 \% C.L.
(inside) and 95 \% C.L. (outside) boundaries constrained by
Planck+WP+BAO+high-$\ell$+BICEP2 with dotted ellipses. 
In the left panel, the three large black points correspond to the
theoretical prediction of the model (i) for $N=50, 55, 60$, whereas
the small black points represent the prediction of the model (ii)
for $f/M_{\rm pl}=5,6,7,8,9,10$ with $N=50, 55, 60$.  In the right
panel, the theoretical curves are plotted for $N=55,58,63$ with
several different values of $\xi$ between $0$ and $-10^{-2}$.}
\end{figure}

For the CMB likelihood analysis, we expand the primordial power 
spectra ${\cal P}_{\cal R}^{\rm prim}$ and ${\cal P}_T^{\rm prim}$ 
as a Taylor series about a pivot wave number $k_*$.
We employ the consistency relation $r=-8n_t$
and set the scalar and tensor runnings to be 0 
in the likelihood analysis. 
The likelihood results are derived 
with the CosmoMC code \cite{cosmomc,Lewis}
by assuming the flat $\Lambda$CDM model and $N_{\rm eff}=3.046$ 
relativistic degrees of freedom with an instant reionization.

In Fig.~\ref{fig2}, we plot the 68 \% C.L. and 95 \% C.L. regions 
(plotted as the gray shaded color)
in the $(n_s,r)$ plane constrained by 
the joint data analysis of Planck \cite{planck}, WP \cite{WMAP9}, 
baryon acoustic oscillations (BAO) \cite{BAO}, 
and Atacama Cosmology Telescope/South Pole Telescope temperature
data of high multipoles $\ell$ (high-$\ell$) \cite{highl}.
The tensor-to-scalar ratio is constrained to be $r<0.15$ at 
95 \% C.L. We have chosen the value $k_*=0.05$ Mpc$^{-1}$ 
in Fig.~\ref{fig2}, but the results are insensitive to the choice 
of $k_*$ as long as the pivot scale is in the observed range of CMB. 
Hence, the likelihood contours for different choices of $k_*$
are similar to those given in Fig.~\ref{fig2}.
We also show the 68 \% C.L. and 95 \% C.L. boundaries (dotted curves) 
constrained by adding the BICEP2 data \cite{bicep2} to the 
Planck+WP+BAO+high-$\ell$ data.
In this case, $r$ is bounded from below.

For the monomial potential $V(\phi)=\lambda \phi^n/n$ with $\xi=0$, 
the observables (\ref{efold}) reduce to \cite{reviews}
\be
n_s=1-\frac{2(n+2)}{4N+n}\,,\qquad
r=\frac{16n}{4N+n}\,.
\label{nsrcha}
\ee
From Eq.~(\ref{efoldq=3d}), the number of e-foldings for the 
quadratic potential ($n=2$) depends on the reheating temperature 
$T_{\rm RH}$ and the wave number $k$.
For the pivot scale $k_*=0.05$ Mpc$^{-1}$ the number 
of e-foldings is in the range $48.7<N (k_*)<53.3$ for 
$10^4~{\rm GeV}<T_{\rm RH}<10^{10}~{\rm GeV}$. 
If $k_*=a_0H_0$, then $54.1<N (k_*)<58.8$ for 
$10^4~{\rm GeV}<T_{\rm RH}<10^{10}~{\rm GeV}$. 
In Fig.~\ref{fig2}, we plot the theoretical curves of the model (i)
in the $(n_s, r)$ plane for $N$ between 50 and 60. 
For $N>52$ the quadratic potential is inside the 95 \% C.L. 
boundary constrained by the 
Planck+WP+BAO+high-$\ell$ data.

In Fig.~\ref{fig2}, we also show the theoretical prediction of 
the model (ii) for $N=50, 55, 60$ with $f$ between 
$5M_{\rm pl}$ and $10M_{\rm pl}$.
When $N>50$, natural inflation is inside the 95 \% C.L. boundary 
constrained the Planck+WP+BAO+high-$\ell$ data for $f>5M_{\rm pl}$. 
In the $f \to \infty$ limit, inflation occurs in the region around the 
potential minimum ($\phi=\pi f$), in which case $n_s$ and $r$ 
approach the values (\ref{nsrcha}) with $n=2$.

In the model (iii) with $|\xi| \ll 1$, the number of e-foldings 
(\ref{efoldq=4}) does not depend on $T_{\rm RH}$.
For given $k_*$, the value of $N$ is fixed, e.g.,
$N=58$ for $k_*=0.05$ Mpc$^{-1}$. 
In the right panel of Fig.~\ref{fig2}, we plot the theoretical 
values of $n_s$ and $r$ for $N=55, 58, 60$ with 
several different values of $\xi$.
For $|\xi|$ of the order of $10^{-3}$, the presence of non-minimal 
couplings allows a possibility of reducing the values of $r$ to be compatible 
with the upper bound derived from the Planck+WP+BAO+high-$\ell$ data. 
In the $|\xi| \to \infty$ limit, we obtain the same values of $n_s$ and $r$ 
as those in the Starobinsky $f(R)$ model \cite{staro}, i.e., $n_s=1-2/N$ and 
$r=12/N^2$ \cite{nonper2}. 

\section{The intensity of the gravitational wave background}
\label{secspe}

In this section, we calculate the spectrum of the gravitational wave background 
generated in large-field inflationary models discussed in Sec.~\ref{secinf}.
From the second-order action (\ref{secondten}), the energy density 
$\rho_{\rm GW}$ of gravitational waves corresponds to the (00) component 
of its energy-momentum tensor $T_{\mu \nu}$, i.e., 
\be
\rho_{\rm GW}=\frac{M_{\rm pl}^2F}{8a^2} 
\left[ (h_{ij}')^2+\left(\partial h_{ij} \right)^2 \right]\,.
\label{rhoGW}
\ee
After the perturbations enter the Hubble radius during the radiation 
or matter era, the WKB solution to Eq.~(\ref{heq2}) is given by 
\be
h_{\bm k}^{\mu} \propto a^{-1} e^{\pm ik \tau}\,,
\label{WKB}
\ee
where we have neglected the variation of $F$ relative to that of $a$. 
Substituting Eq.~(\ref{hij}) with the solution (\ref{WKB}) into 
Eq.~(\ref{rhoGW}) and taking the 
spatial average, the energy density reads
\be
\rho_{\rm GW}=\frac{M_{\rm pl}^2F}{2} 
\int \frac{d^3 k}{(2\pi)^3} \frac{k^2}{a^2} \sum_{\mu=+,\times}
|h_{\bm k}^{\mu}|^2\,.
\label{rhoGW2}
\ee
We define the intensity of the gravitational wave 
background as 
\be
\Omega_{\rm GW} \equiv \frac{1}{\rho_c}
\frac{d \rho_{\rm GW}}{d \ln k}=
\frac{1}{12} \left( \frac{k}{aH} \right)^2 
{\cal P}_T\,,
\label{OmeGW}
\ee
where $\rho_c \equiv 3FH^2 M_{\rm pl}^2$ is 
the critical density of the Universe, and
${\cal P}_T$ is the tensor spectrum defined in Eq.~(\ref{PTdef}).
In the second equality of Eq.~(\ref{OmeGW}), we have 
substituted the energy density (\ref{rhoGW2}).

In order to compute the spectrum (\ref{OmeGW}) today, 
we need to solve Eq.~(\ref{heq2}) from the onset of inflation to 
the present epoch. As we have already studied in Sec.~\ref{secinf}, 
the background dynamics during inflation and reheating are 
known by solving Eqs.~(\ref{rhoeq})-(\ref{ddotphi}) numerically. 
To discuss the cosmological dynamics after the radiation era, 
we need to take into account the contribution of relativistic particles, 
non-relativistic particles (dark matter and baryons), and dark energy 
to the Friedmann equation.

The entropy density of relativistic particles at temperature $T$ is 
given by $s(T)=(2\pi^2/45)g_{s} (T) T^3$.
Due to the entropy conservation $sa^3=$\,constant, 
the radiation energy density $\rho_r(T)=(\pi^2/30)g_*(T) T^4$ 
evolves as $\rho_r \propto g_* g_{s}^{-4/3}a^{-4}$. 
The explicit forms of $g_*(T)$ and $g_{s}(T)$ are 
given in Ref.~\cite{kuroyanagi1}. 
The energy density of non-relativistic particles decreases as 
$\rho_m \propto a^{-3}$.
For dark energy, we assume that its energy density $\rho_{\rm DE}$ 
is given by the cosmological constant $\Lambda$. 
We can consider some other sources for dark energy, 
but it hardly affects the resulting gravitational wave spectrum unless
Eq.~(\ref{heq2}) is subject to change under some 
modification of gravity. The term $F$ is very close to 1 
around today, so the effect of non-minimal couplings is negligibly 
small at the late cosmological epoch.

Defining today's density parameters as
$\Omega_{j0}=\rho_{j0}/(3F_0H_0^2M_{\rm pl}^2)$, 
where $j=r,m,\Lambda$ and the subscript ``0'' represents 
the present values, the Friedmann equation after the end 
of reheating can be expressed as 
\be
\frac{H^2}{H_0^2}=\frac{F_0}{F} 
\left[ \left( \frac{g_*}{g_{*0}} \right) \left( 
\frac{g_{s}}{g_{s0}} \right)^{-4/3} \Omega_{r0}
\left( \frac{a}{a_0} \right)^{-4}+
\Omega_{m0} \left( \frac{a}{a_0} \right)^{-3}+
\Omega_{\Lambda 0} \right]\,.
\label{hubble}
\ee
For numerical simulations, we take the present radiation density
to be $\Omega_{r0} h^2=4.15\times 10^{-5}$ and use the mean
likelihood values derived from the Planck+WP+BAO+high-$\ell$ 
data \cite{planck}: the non-relativistic matter density $\Omega_{m0} h^2=0.141$, 
the dark energy density $\Omega_{\Lambda 0}=0.692$, the amplitude of primordial
curvature perturbations ${\cal P}_{\cal R}^{\rm prim}=2.2\times
10^{-9}$, and the normalized Hubble constant $h=0.678$.  For the
calculation of relativistic degrees of freedom, we only include
particles in the standard model of particle physics, where the maximum
values of $g_*$ and $g_{s}$ are $106.75$.  

Since the primordial tensor perturbation is frozen by the second 
horizon crossing characterized by $k=a_{\rm hc}H_{\rm hc}$, 
today's power spectrum ${\cal P}_{T0}$
is related to the primordial one ${\cal P}_T^{\rm prim}$ via
${\cal P}_{T0}={\cal P}_T^{\rm prim} (a_{\rm hc}/a_0)^2$.
If the scale factor evolves as $a \propto t^p$, where $p$ is 
a constant, the Hubble parameter $H$ is proportional to 
$a^{-1/p}$, so that $a_{\rm hc} \propto k^{p/(p-1)}$. 
Then, today's gravitational wave intensity (\ref{OmeGW}) 
has the momentum dependence
\be
\Omega_{\rm GW} \propto k^{n_t+2(2p-1)/(p-1)}\,,
\ee
where $n_t$ is the primordial tensor spectral index. 

During the radiation era ($p=1/2$) and the matter era ($p=2/3$), 
we have $\Omega_{\rm GW} \propto k^{n_t}$ and 
$\Omega_{\rm GW} \propto k^{n_t-2}$, respectively. 
The nearly scale-invariant property of the primordial tensor 
perturbation is imprinted for the modes that re-entered the
Hubble radius during the radiation-dominated epoch.
This is the case for the frequencies $f=k/(2\pi)$ associated 
with the direct detection of gravitational waves by DECIGO 
or BBO ($\sim 1$ Hz). For the perturbations that re-entered the
Hubble radius during the matter era, the intensity 
$\Omega_{\rm GW}$ has a highly red-tilted spectrum.

For the inflationary models like (i) and (ii), the reheating stage 
corresponds to a temporal matter-dominated epoch 
driven by a massive inflation field.
Then, there is a suppression of $\Omega_{\rm GW} $ at high frequencies. 
This suppression occurs for \cite{Kamionkowski}
\be
f>f_{\rm RH} \equiv 0.26\left(\frac{T_{\rm RH}}{10^7\,{\rm GeV}}\right)
\left(\frac{g_{*,{\rm RH}}}{100}\right)^{1/2}
\left(\frac{g_{s,{\rm RH}}}{100}\right)^{-1/3} {\rm Hz}\,.
\label{freq}
\ee
For increasing $\Gamma$, the critical frequency $f_{\rm RH}$ 
becomes larger. In the model (iii) with $|\xi| \ll 1$, 
the evolution of the scale factor during reheating is given by 
$a \propto t^{1/2}$, so the suppression of $\Omega_{\rm GW}$ 
mentioned above is absent. 
As we explained in Sec.~\ref{secback}, we compute the reheating temperature 
numerically at the time when the field energy 
density $\rho_{\phi}$ drops below the radiation 
energy density $\rho_r$.

\begin{figure}
\begin{center}
\includegraphics[height=2.7in,width=3.35in]{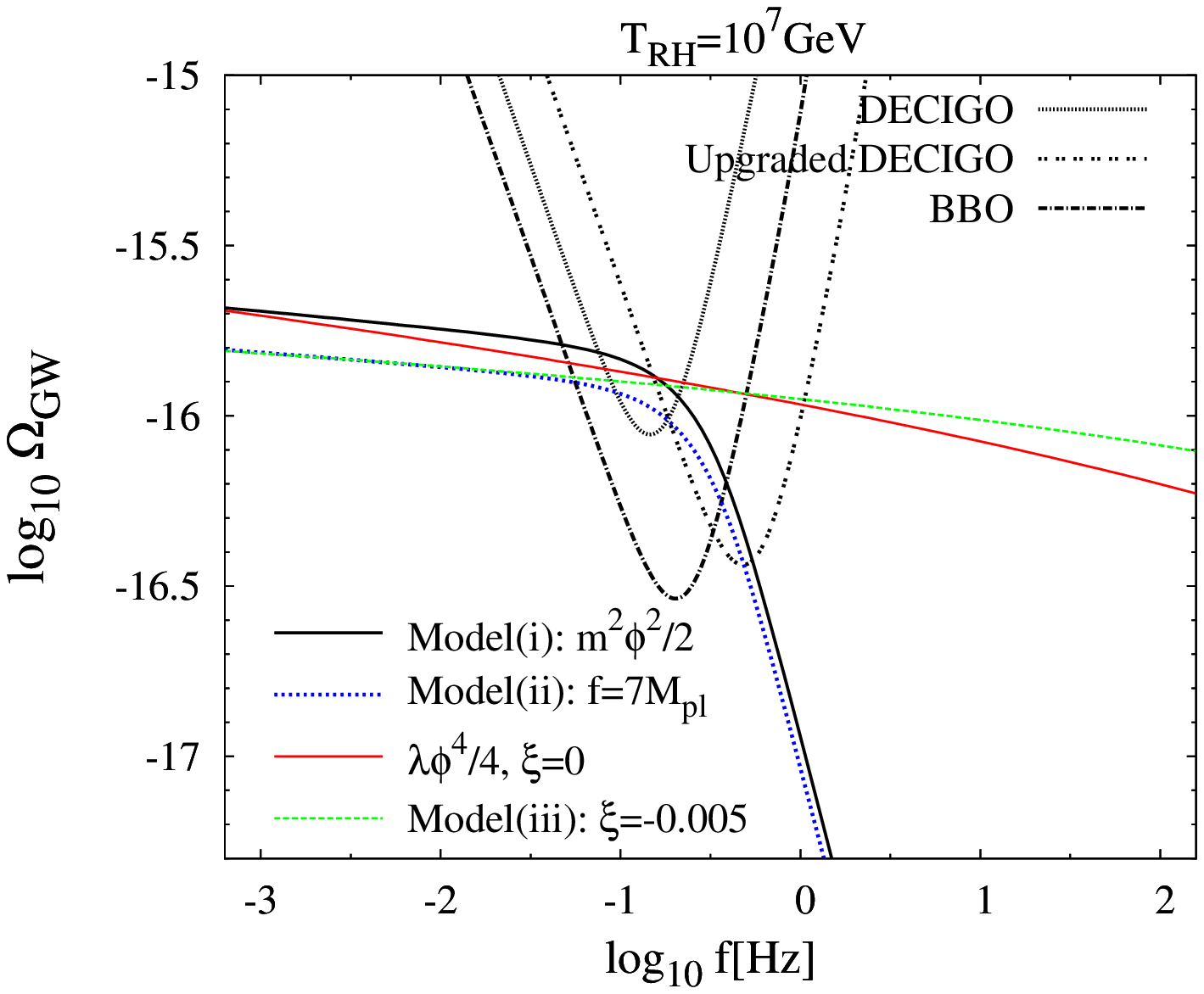}
\includegraphics[height=2.7in,width=3.6in]{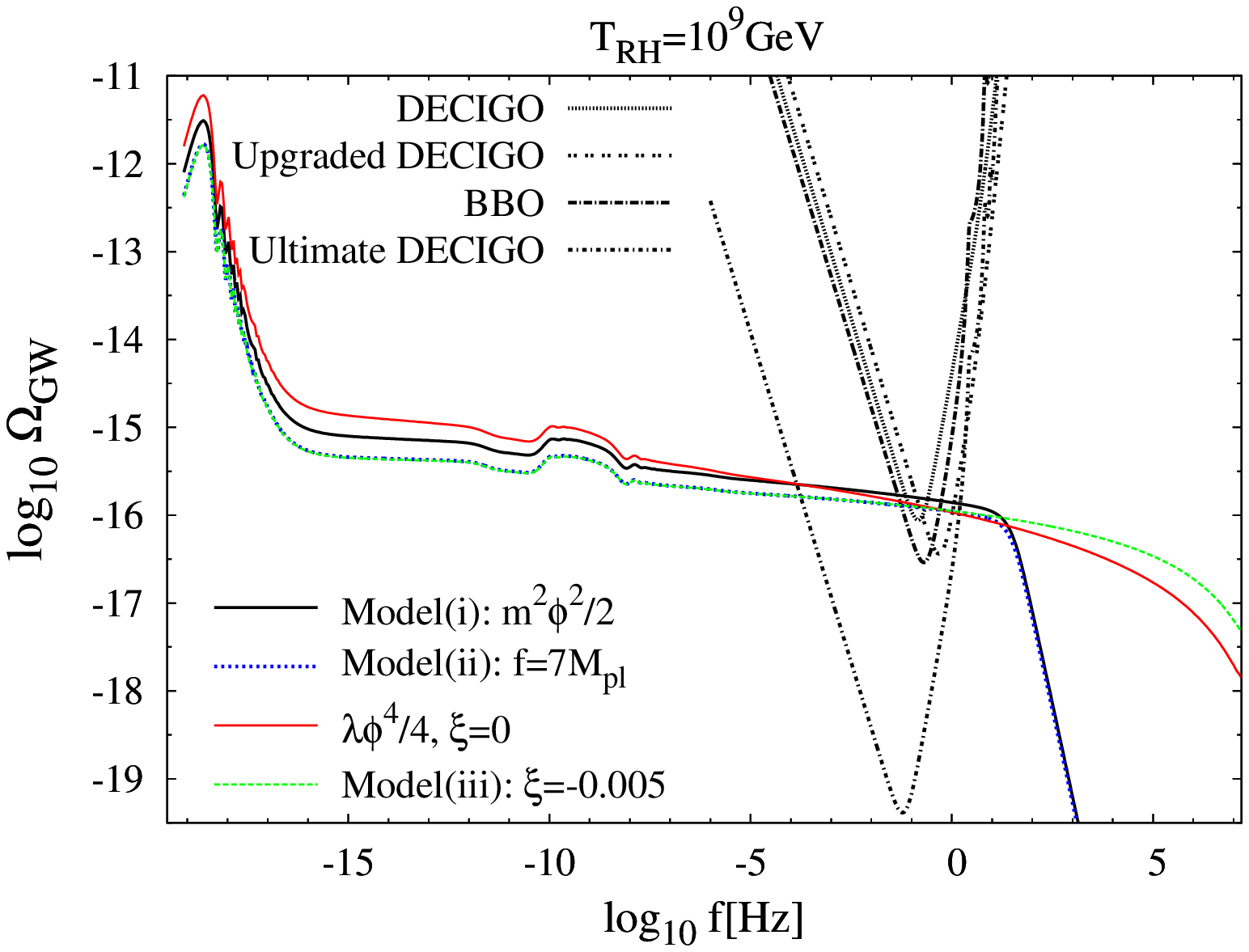}
\caption{\label{fig3} Today's intensity $\Omega_{\rm GW}$ of the
gravitational wave background versus the frequency $f=k/(2\pi)$ for
the four models: (i) $V(\phi)=m^2 \phi^2/2$ with $\xi=0$, (ii)
$V(\phi)=\Lambda^4 [1+\cos (\phi/f)]$ with $\xi=0$ and $f=7M_{\rm pl}$, 
the potential $V(\phi)=\lambda \phi^4/4$ with $\xi=0$,
and (iii) $V(\phi)=\lambda \phi^4/4$ and $\xi=-5.0
\times 10^{-3}$.  We also show the sensitivity curves for DECIGO,
upgraded DECIGO, BBO, and Ultimate DECIGO.
Each panel corresponds to the spectra
around the detection sensitivity with $T_{\rm RH}=10^7$~GeV (left)
and the spectra in the wide range of frequencies with $T_{\rm
RH}=10^9$~GeV (right).}
\end{center}
\end{figure}

Under the slow-roll approximation the primordial spectral index is 
given by $n_t=-2\epsilon_U$.  
Note, however, that this formula 
overestimates the amplitude of the gravitational wave 
spectrum at $f \sim 1$ Hz  by 20\,\% for chaotic inflation 
models \cite{kuroyanagi1}.   
Therefore, we compute the intensity $\Omega_{\rm GW}$ for 
three inflationary models (i)-(iii) without employing the 
slow-roll approximation. We numerically solve Eq.~(\ref{heq2}) 
for each wave number $k$ together with the background equations 
of motion (\ref{rhoeq})-(\ref{ddotphi}) during inflation/reheating 
and Eq.~(\ref{hubble}) after the end of reheating.

In Fig.~\ref{fig3}, we show today's gravitational wave intensity
versus the frequency $f$ for the three large-field inflationary models
(i)-(iii) with $T_{\rm RH}=10^7$~GeV (left) and $T_{\rm RH}=10^9$~GeV
(right).  Recall that the reheating temperatures for the models (i)
and (ii) are related to the number of e-foldings $N$, as shown in
Fig.~\ref{fig1}.  In the left panel, we see that the anisotropic stress
due to neutrino free streaming \cite{watanabe,weinberg,kuroyanagi1} at
$f<10^{-9}$\,Hz does not affect the amplitude of the gravitational
waves at the direct detection scales ($\sim 10^{-1}$Hz).  Since the primordial
tensor-to-scalar ratio in the model (i) is greater than that in the
model (ii) (see Fig.~\ref{fig2}), the amplitude of $\Omega_{\rm GW}$
in the former is larger than that in the latter at the CMB scale
($\sim 10^{-18}$Hz).  In Fig.~\ref{fig3}, the suppression of
$\Omega_{\rm GW}$ due to the presence of the temporal matter era after
inflation is clearly seen for large frequencies. As estimated by
Eq.~(\ref{freq}), the critical frequency $f_{\rm RH}$ becomes larger
for increasing $T_{\rm RH}$.

In the model (iii) with $|\xi| \ll 1$, the reheating stage is characterized 
by the oscillation of a massless inflaton, so the suppression of 
$\Omega_{\rm GW}$ mentioned above is not present. 
However, due to the steepness of the potential, the evolution of 
the inflaton is faster than that in the models (i) and (ii) 
around the end of inflation.
As we see in the right panel of Fig.~\ref{fig3}, this
leads to the mild decrease of the power spectrum 
around the frequencies $f \gtrsim 10^4$~Hz.

In Fig.~\ref{fig3}, we also plot the sensitivity curves for DECIGO, 
upgraded DECIGO, BBO, and Ultimate DECIGO.
In particular, the sensitivity of 
upgraded DECIGO is improved over DECIGO to cover 
a wider range of frequencies.
In the model (i), BBO, and upgraded DECIGO can 
potentially measure inflationary gravitational waves 
for $T_{\rm RH} \gtrsim 10^6$ GeV and  
$T_{\rm RH} \gtrsim 10^7$ GeV, respectively.
In the model (ii), the intensity $\Omega_{\rm GW}$ becomes 
smaller for decreasing $f$, but the detection is still possible 
for $f \gtrsim 4M_{\rm pl}$.
The model (iii) allows the possible detection in the frequency range 
$0.1~{\rm Hz} \lesssim f \lesssim 1~{\rm Hz}$, irrespective 
of the reheating temperature.
In the next section, we shall discuss this issue in more detail
by taking into account the noise associated with the 
interferometric detectors.

\section{Detectability of inflationary gravitational waves}
\label{secsnr}

In order to discuss the detectability of inflationary gravitational waves, 
we compute the signal-to-noise ratio (SNR) by two interferometric
detectors. Here the SNR is expressed in terms of $\Omega_{\rm GW}(f)$ 
given by Eq.~(\ref{OmeGW}) and the functions related to the
experimental design, such as the noise spectrum $N_{I,J}(f)$ and 
the overlap reduction function $\gamma_{IJ}(f)$, as \cite{Allen}
\be
[{\rm SNR}]^2=2\left(\frac{3H_0^2}{10\pi^2}\right)^2T_{\rm obs}
\sum_{(I,J)}\int^{\infty}_0df\frac{|\gamma_{IJ}(f)|^2
\Omega_{\rm GW}^2(f)}{f^6N_I(f)N_J(f)},
\ee
where $T_{\rm obs}$ is the duration of observations, and the
subscripts $I$ and $J$ refer to independent signals obtained at
each detector, or observables generated by combining the detector
signals. The overlap reduction function $\gamma_{IJ}(f)$ can 
be calculated with information about relative locations and 
orientations of detectors \cite{Kudoh,Corbin}. 
In the following, we present explicit forms of the noise spectra 
for DECIGO, upgraded DECIGO, BBO, and Ultimate DECIGO.

\begin{itemize}
\item DECIGO

DECIGO is planned to be a Fabry-Perot Michelson interferometer 
with an arm length of $L=1.0\times 10^6$~m \cite{decigo}. 
The noise spectral density is given by \cite{Kudoh} 
\be
N_1(f)=N_2(f)=S_{\rm shot}^2+S_{\rm accel}^2+S_{\rm rad}^2\,,
\ee
where $S_{\rm shot}=2.3\times10^{-24}\sqrt{(1+f^2/f_c^2)}\,{\rm Hz}^{-1/2}$ is the shot noise 
\footnote{Here we have fixed a typo in Ref.~\cite{Kudoh} as already done in \cite{himemoto,kuroyanagi}.}, 
$S_{\rm accel}=2.0\times 10^{-26}/(f/{\rm Hz})^2\,{\rm Hz}^{-1/2}$ is the acceleration noise, and
$S_{\rm rad}=6.0 \times 10^{-26}/[(f/{\rm Hz})^2 \sqrt{1+f^2/f_c^2}]\,{\rm
  Hz}^{-1/2}$ is the radiation pressure noise, with $f_c=7.36$Hz being
the cutoff frequency.

\item Upgraded DECIGO

In order to resolve all foreground gravitational waves coming from
neutron star binaries, it is necessary to improve the sensitivity
of DECIGO by a factor of 3 \cite{Yagi:2011wg}. We consider the following noise
spectrum as an example of an upgraded version of 
DECIGO: $S_{\rm shot}=3.3\times10^{-25}\sqrt{(1+f^2/f_c^2)}\,{\rm Hz}^{-1/2}$ 
and $S_{\rm accel}=8.0\times 10^{-26}/(f/{\rm Hz})^2\,{\rm Hz}^{-1/2}$ with the 
cutoff frequency $f_c=3.57$\,Hz, which would be possible
by upgrading the arm length from $10^6$\,m to $1.5\times 10^6$\,m,
the laser power from $10$\,W to $30$\,W and mirror radius from 
$0.5$\,m to $0.75$\,m. 
For radiation pressure, we adopt the same noise spectrum 
as that of DECIGO, which is negligible compared to the
acceleration noise.  

\item BBO

BBO adopts a technique called Time-Delay Interferometry (TDI), 
in which new variables ($I=A,E,T$) are constructed to cancel 
the laser frequency noise. The noise transfer functions for 
the TDI variables are given by \cite{Prince}
\ba
& &
N_A(f)=N_E(f)=8\sin^2(\hat{f}/2)
\left\{ (2+\cos\hat{f})N_{\rm shot}
+2[3+2\cos\hat{f}+\cos(2\hat{f})]N_{\rm accel} \right\}\,,\\
& &
N_T(f)=2(1+2\cos\hat{f})^2
[N_{\rm shot}+4\sin^2(\hat{f}/2)N_{\rm accel}]\,,
\ea
where $\hat{f}=2\pi Lf$. 
In the case of BBO, the arm length is $L=5.0\times 10^7$\,m and 
the noise functions are $N_{\rm shot}=2.0\times 10^{-34}/(L/{\rm m})^2~{\rm Hz}^{-1}$ 
and $N_{\rm accel}=9.0\times 10^{-34}/[(2\pi f/{\rm Hz})^4 \cdot (2L/{\rm m})^2]~{\rm Hz}^{-1}$.

\item Ultimate DECIGO

Ultimate DECIGO is an ideal experiment whose noise is limited only by quantum noise \cite{decigo}. 
We consider a TDI type experiment whose arm length is $L=5.0\times 10^7$\,m and 
the noise functions are $N_{\rm shot}=2.72\times 10^{-36}/(L/{\rm m})^2~{\rm Hz}^{-1}$ 
and $N_{\rm accel}=9.0\times 10^{-38}/[(2\pi f/{\rm Hz})^4 
\cdot (2L/{\rm m})^2]~{\rm Hz}^{-1}$ \cite{Kudoh}.\footnote{Note that the shot noise 
is corrected from that of Ref.~\cite{Kudoh}.}

\end{itemize}

\begin{figure}
\begin{center}
\includegraphics[height=2.2in]{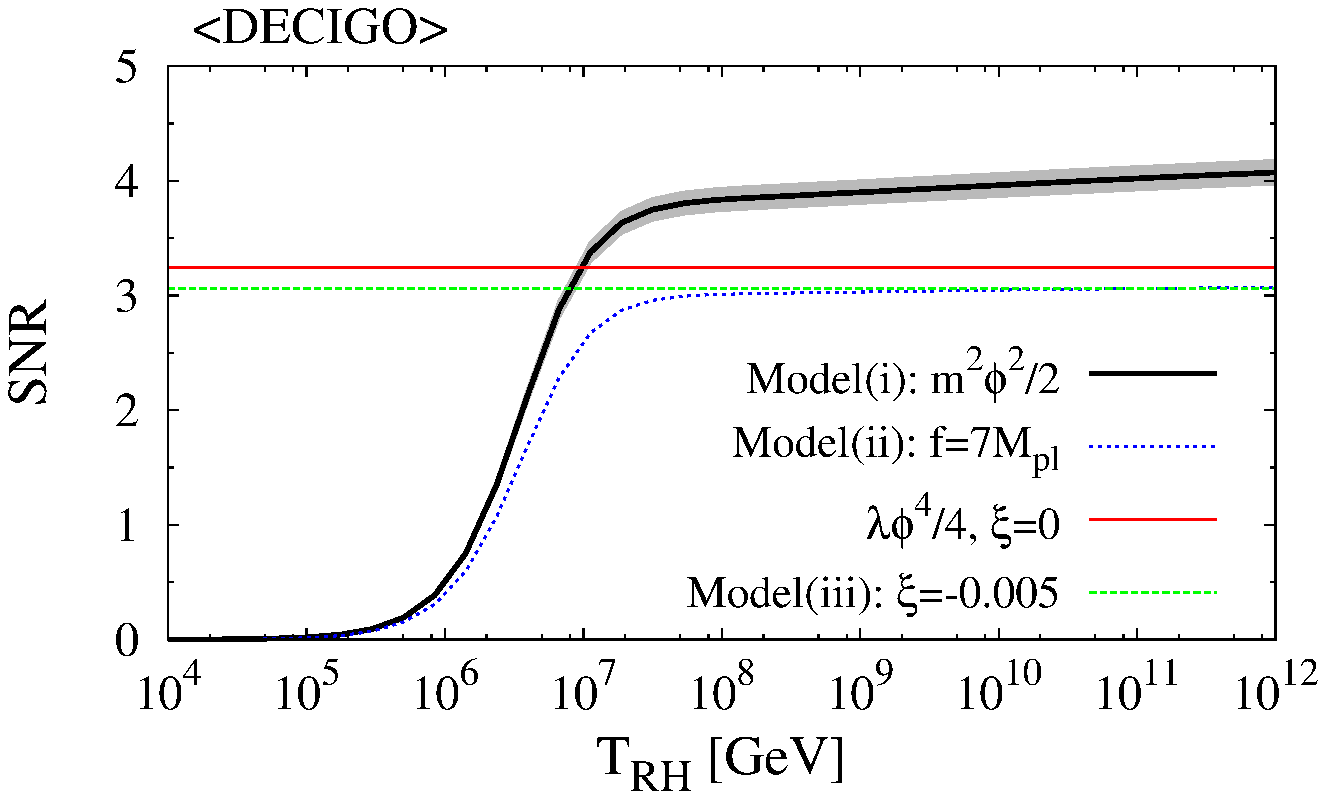}\\
\includegraphics[height=2.2in]{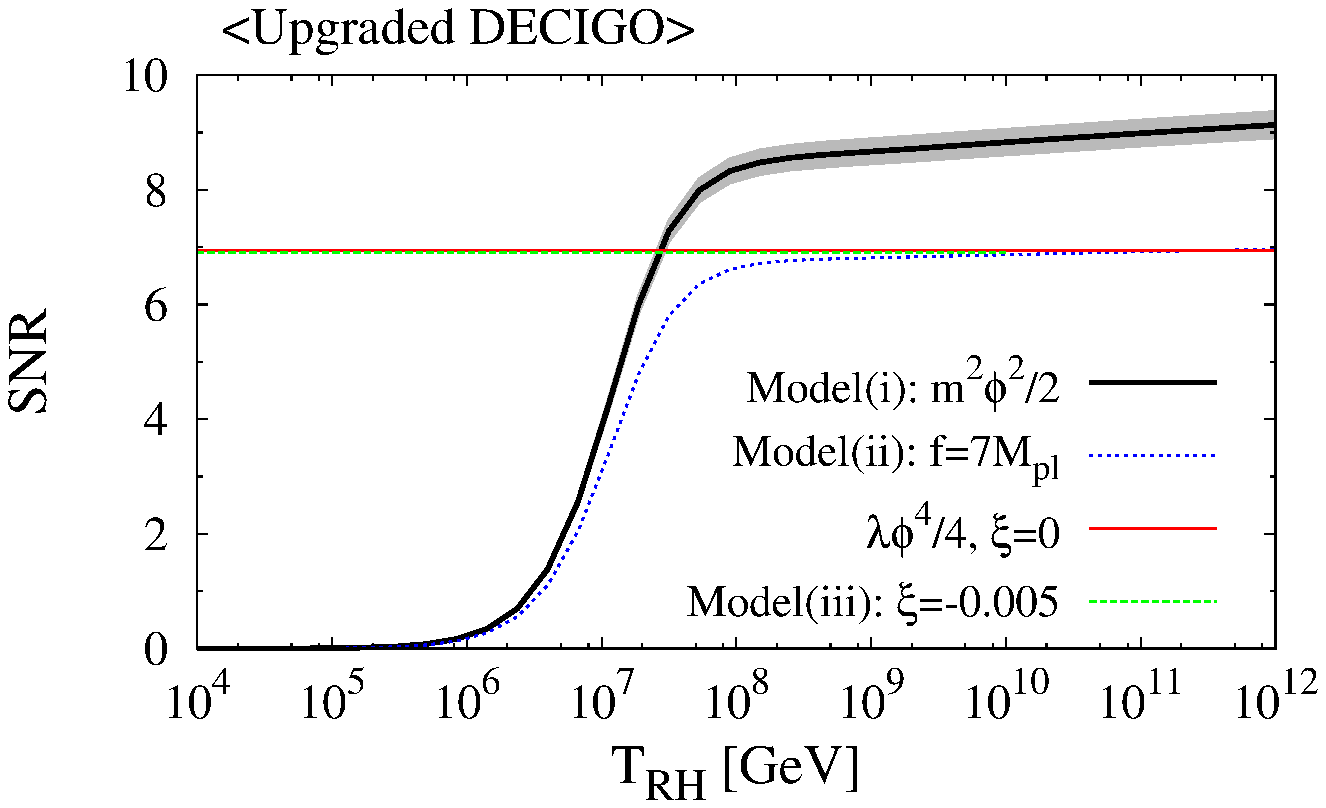}\\
\includegraphics[height=2.2in]{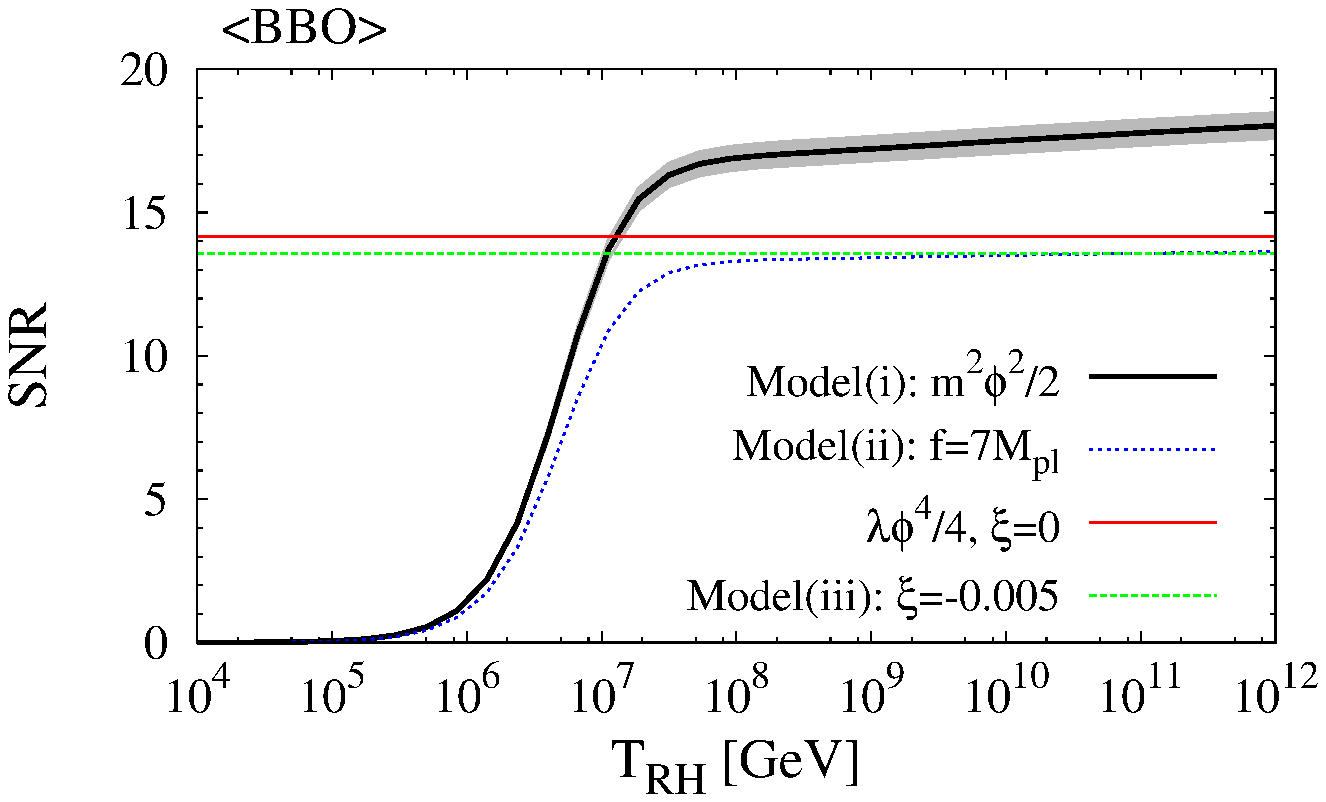}\\
\includegraphics[height=2.2in]{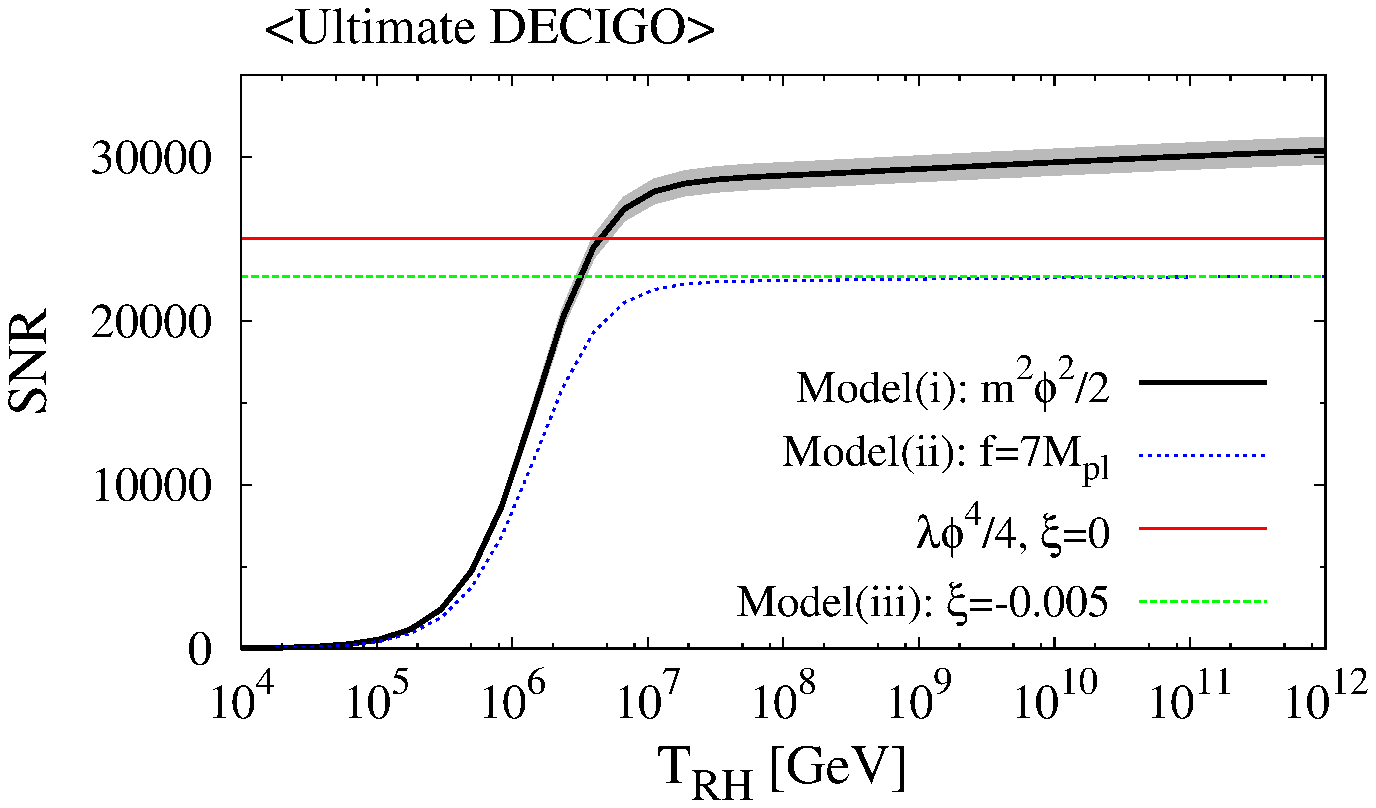}
\caption{\label{SNR1} The SNR versus the reheating temperature
computed for the model (i), the model (ii) with $f=7M_{\rm pl}$,
the potential $V(\phi)=\lambda \phi^4/4$ with $\xi=0$, and
the model (iii) with $\xi=-5.0 \times 10^{-3}$.  Each
panel corresponds to the SNR for 
DECIGO, upgraded DECIGO, BBO, and Ultimate DECIGO, from top to
bottom, respectively. The gray region shows the 1$\sigma$
uncertainty in the amplitude of primordial curvature perturbations
${\cal P}_{\cal R}^{\rm prim} =(2.200^{+0.056}_{-0.054})\times
10^{-9}$ constrained from Planck \cite{planck}.  For observation time
of the gravitational wave, we assume $T_{\rm obs}=10~{\rm year}$.  The
SNR scales as $\propto \sqrt{T_{\rm obs}/10~{\rm year}}$. }
\end{center}
\end{figure}

\begin{figure}
\begin{center}
\includegraphics[height=2.45in,width=3.5in]{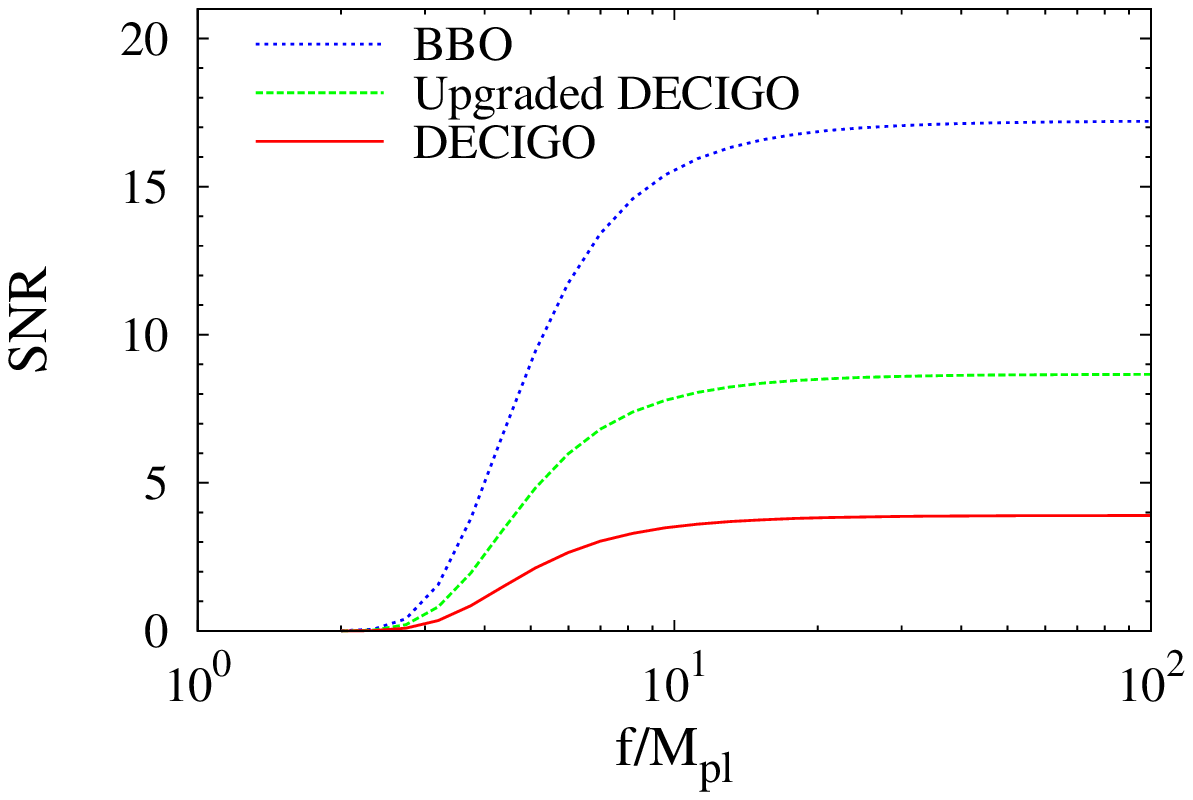}
\includegraphics[height=2.45in,width=3.5in]{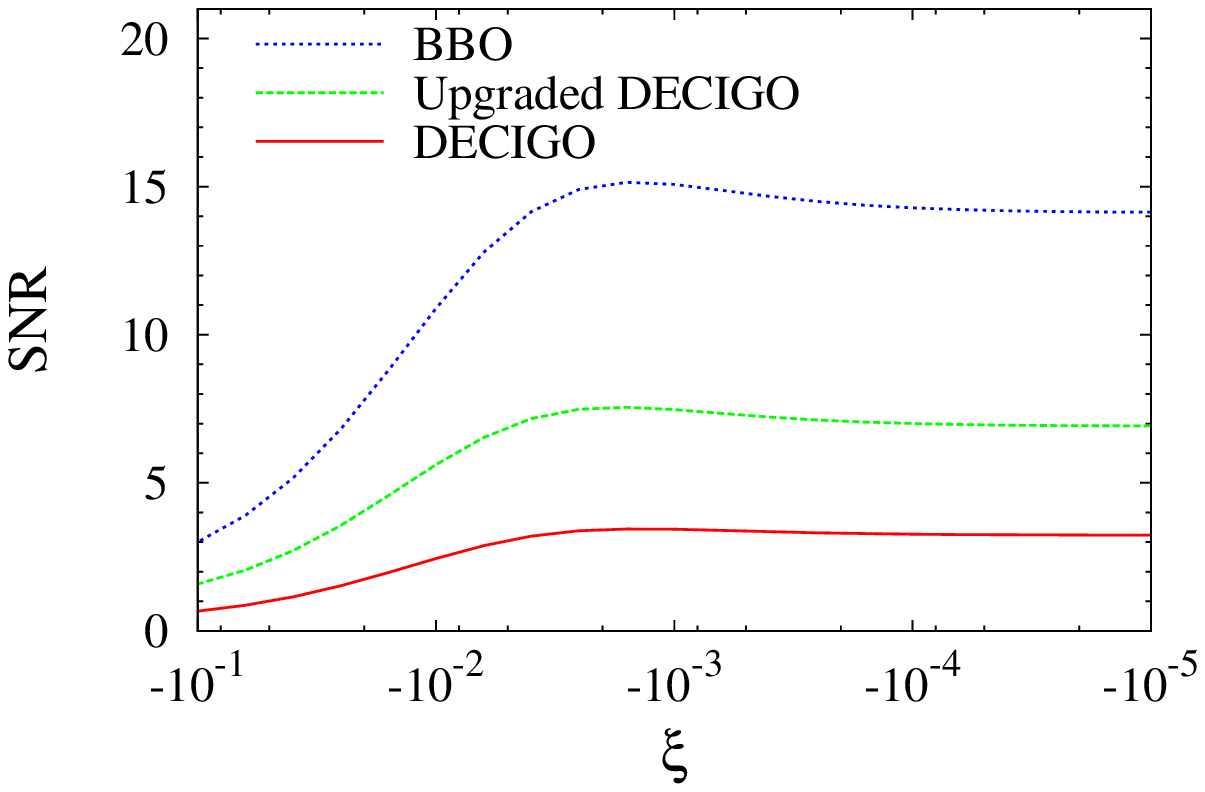}
\caption{\label{SNR2} 
(Left) Dependence of the SNR on the parameter 
$f/M_{\rm pl}$ for the model (ii). 
(Right) Dependence of the SNR on the non-minimal 
coupling $\xi$ for the model (iii).
Each line is the SNR calculated with noise curves of 
DECIGO, upgraded DECIGO and BBO, 
respectively, with $T_{\rm obs}=10$~year.  
In both cases, we assume $T_{\rm RH}=10^9$~GeV.  }
\end{center}
\end{figure}

In Fig.~\ref{SNR1}, we show the SNR versus the reheating temperature
$T_{\rm RH}$ for the cross-correlation analysis expected with 10-year
observations by DECIGO, upgraded DECIGO, 
BBO, and Ultimate DECIGO.\footnote{These SNRs are slightly larger
than those computed in Ref.~\cite{kuroyanagi} by a factor
$(220/106.75)^{1/3} \simeq 1.3$ because we consider the standard
model particles alone and take $g_{*,{\rm RH}}=106.75$ rather than
$g_{*,{\rm RH}}=220$ used in Ref.~\cite{kuroyanagi}.}
For the models (i) and (ii), the SNR decreases significantly for 
$T_{\rm RH}$ lower than $10^7$~GeV because of the
suppression due to the presence of the temporal matter era 
after inflation. The SNR of the model (ii) is smaller than that 
of the model (i) for the same reheating temperature.
As we see in the left panel of Fig.~\ref{SNR2}, the SNR of 
natural inflation increases for larger $f$ to approach that 
of the model (i) in the $f \to \infty$ limit. 
This tendency is also seen in Table \ref{table}, in 
which the SNR as well as $r$ and $\Omega_{\rm GW}$ 
are given for $T_{\rm RH}=10^{9}$~GeV.

In Fig.~\ref{SNR1}, we find that the SNR of the model (iii) is practically 
independent of the reheating temperature for $|\xi| \ll 1$. 
The primordial tensor-to-scalar ratio $r$ gets smaller for increasing 
$|\xi|$. As we see in the right panel of Fig.~\ref{SNR2}, the SNR 
starts to decrease for $\xi \lesssim -2 \times 10^{-3}$.
This is inside the 95\,\%\,C.L. region constrained by 
the Planck+WP+BAO+high-$\ell$ data (see Fig.~\ref{fig2}). 

If the SNR is larger than 3, the primordial gravitational waves could be
directly measured at $3\sigma$ (under the assumption of Gaussian noise).  
For the model (i), the detections by
DECIGO, upgraded DECIGO, BBO, and Ultimate DECIGO 
at $3\sigma$ would be possible if
the reheating temperature is larger than 
$7.8 \times 10^6$~GeV, $7.9 \times 10^6$~GeV, 
$1.8 \times 10^6$~GeV, and $2.8 \times 10^3$~GeV,
respectively.  
The detections by upgraded DECIGO (BBO) are also feasible 
at 5$\sigma$ if $T_{\rm RH}>1.5\times 10^7$ GeV ($2.8 \times 10^6$~GeV).

The situation in the model (ii) is similar to that 
in the model (i). For $f=7M_{\rm pl}$, DECIGO, upgraded DECIGO 
and BBO could measure the primordial gravitational waves at 
$3\sigma$ if $T_{\rm RH}>6.8\times 10^7$~GeV, 
$T_{\rm  RH}>9.9\times 10^6$~GeV, 
$T_{\rm RH}>2.2 \times 10^6$~GeV, and $T_{\rm RH}>3.3 \times 10^3$~GeV,
respectively.  The detections would be possible at 5$\sigma$  for
$T_{\rm RH}>2.2\times 10^7$ GeV (upgraded DECIGO)  
and $T_{\rm RH}>2.2\times 10^7$ GeV (BBO).  
The SNR increases for larger $f$ and converges to that of the model (i). 
For $T_{\rm RH}>10^8$ GeV
the SNR is insensitive to the reheating temperature.
The detections by upgraded DECIGO (BBO) would be
possible for  $f>4.2 M_{\rm pl}$ ($f>3.6 M_{\rm pl}$) at $3\sigma$ 
and for $f>5.3 M_{\rm pl}$ ($f>4.0 M_{\rm pl}$) at 5$\sigma$. The $f$ 
dependence of the SNR for Ultimate DECIGO is the same as that of DECIGO/BBO
seen in the left panel of Fig.~\ref{SNR2}, but the scale of the vertical axis is much 
larger because of the extremely high sensitivity.  Thus, Ultimate DECIGO can 
detect the signal with a very good accuracy as long as we consider
the parameter region indicated by Planck, $f\gtrsim 5 M_{\rm pl}$. 

The SNR in the model (iii) is independent of $T_{\rm RH}$, 
but it depends on non-minimal couplings $\xi$.
From the right panel of Fig.~\ref{SNR2}, we find that 
the detections at $3\sigma$ by DECIGO, upgraded DECIGO and BBO 
would be feasible for $\xi >-5.4\times 10^{-3}$, $\xi>-0.035$ and $\xi>-0.1$, 
respectively. 
The 5$\sigma$ detections by upgraded DECIGO  (BBO) would 
require $\xi>-0.013$ ($-0.043$). 
This corresponds to the regime in which the 
tensor-to-scalar ratio is larger than 0.05.
If the B-mode polarization measurements convincingly 
confirm the existence of primordial gravitational waves 
with $r$ larger than 0.05, its direct detection by upgraded 
DECIGO and BBO could be also possible in the future.

\begin{table}
\begin{center}
\begin{tabular*}{1.0\textwidth}{@{\extracolsep{\fill}}lccccc}
\hline
\hline
\rule[-3pt]{0pt}{12pt}Model & $r$ & $\Omega_{\rm GW}$ & SNR (DECIGO) & 
SNR (Upgraded DECIGO) & SNR (BBO) \\
\hline
\rule[0pt]{0pt}{10pt}
Model (i)  & 0.153 & $1.55\times 10^{-16}$ & 3.90 & 8.67 & 17.2  \\
Model (ii) ($f=10 M_{\rm Pl}$) & 0.117 & $1.40\times 10^{-16}$ & 3.52 & 7.87 & 15.6  \\
Model (ii) ($f=7 M_{\rm Pl}$) & 0.086 & $1.21\times 10^{-16}$ & 3.03  & 6.81 & 13.4  \\
$V(\phi)=\lambda\phi^4/4$ ($\xi=0$)  & 0.275 & $1.27\times 10^{-16}$ & 3.24 & 6.94 & 14.2  \\
Model (iii) ($\xi=-0.002$) & 0.144 & $1.36\times 10^{-16}$ & 3.43 & 7.56 & 15.1  \\
Model (iii) ($\xi=-0.005$) & 0.085 & $1.22\times 10^{-16}$ & 3.06 & 6.90 & 13.6  \\
\hline
\hline
\end{tabular*}
\caption{The tensor-to-scalar ratio $r$ for the pivot wave number 
$k_*=0.05$~Mpc$^{-1}$, the intensity of the gravitational wave 
$\Omega_{\rm GW}$ at the frequency $f=0.2$~Hz, 
and the SNR in DECIGO, upgraded DECIGO, and BBO
for the three inflationary models listed in (\ref{mo1})-(\ref{mo3}). 
The reheating temperature is set as 
$T_{\rm RH}=10^{9}$~GeV. \label{table}}
\end{center}
\end{table}

\section{Summary} 
\label{secsum}

In this paper, we have discussed the possibility of the direct measurement of 
primordial gravitational waves by next generation interferometric detectors.
The prospects for the direct detection depend on the ratio $r$ between tensor 
and scalar perturbations generated during inflation. 
If the B-mode polarization measurements of the CMB really confirm 
the presence of primordial tensor modes with $r$ of the 
order of 0.1, it is expected that the interferometric detectors like 
DECIGO and BBO will be able to detect inflationary gravitational waves 
in the future.

We have focused on the three large-field inflationary models:  
quadratic chaotic, natural, and quartic chaotic with non-minimal couplings. 
The tensor-to-scalar ratio in the model (i) is larger than that 
in the model (ii) for a given number of e-foldings during inflation. 
We have studied the background dynamics 
after inflation including the discussion of the inflaton decay 
associated with the reheating temperature.
The model (iii) with $|\xi| \ll 1$ can be distinguished from 
the models (i) and (ii), in that the reheating dynamics in 
the former is driven by a massless inflaton field 
rather than the massive field.
This difference affects the resulting spectrum of
the gravitational wave background. 
We have also precisely estimated the number of 
e-foldings relevant to the CMB anisotropies, 
which is important to place accurate constraints 
on inflationary models from the CMB.

We illustrated theoretical predictions (Sec.~\ref{secinf})
for the CMB observables in each inflationary model 
as well as the likelihood contours constrained from 
the joint data analysis of Planck+WP+BAO+high-$\ell$ 
without employing the slow-roll approximation. 
Adding the recent BICEP2 data in the analysis leads to 
the bound: $0.07<r<0.25$ (95\,\% C.L.).
Since there is a tension between the Planck and BICEP2 data,
we have not literally used the constraints derived from 
the BICEP2 data. The future independent B-mode polarization 
measurements will provide us with more convincing 
bounds on $r$.

We have computed the spectra of the gravitational wave background 
$\Omega_{\rm GW}$ for the parameter space 
inside the 95\,\% C.L. boundary constrained by 
Planck+WP+BAO+high-$\ell$. 
For the models (i) and (ii) there is a suppression of 
$\Omega_{\rm GW}$ for the frequencies satisfying (\ref{freq}). 
In this case, the critical frequency $f_{\rm RH}$ depends on the reheating 
temperature $T_{\rm RH}$. Provided that $T_{\rm RH}$ is larger than 
the order of $10^6$ GeV, the models (i) and (ii) can reach the 
detection sensitivity of gravitational waves by 
DECIGO and BBO for $0.1\,{\rm Hz}<f<1\,{\rm Hz}$
(see Fig.~\ref{SNR1}). 
When $|\xi| \ll 1$, the model (iii) is not plagued by the sharp 
suppression of $\Omega_{\rm GW}$ due to the absence 
of a temporal matter era after inflation, so the direct detection 
is possible regardless of the value of $T_{\rm RH}$.

We have calculated the signal-to-noise ratio of DECIGO, upgraded
DECIGO, BBO, and Ultimate DECIGO
for the models (i)-(iii).  Compared to the SNR of
DECIGO computed in Ref.~\cite{kuroyanagi}, we considered the upgraded
version of DECIGO that improves the sensitivity of measurements.  For
the model (i) the direct detections by upgraded DECIGO and BBO are
possible for $T_{\rm RH}>1.5\times 10^7$~GeV, 
$T_{\rm RH}>2.8 \times 10^6$~GeV, and $T_{\rm RH}>4.0 \times 10^3$~GeV 
at $5\sigma$, 
respectively.  For the model (ii) the
upgraded DECIGO and BBO could detect the primordial gravitational waves
at 5$\sigma$ for $f>5.3 M_{\rm pl}$ and $f>4.0 M_{\rm pl}$, respectively, 
provided that $T_{\rm RH}\simg10^8$~GeV.  For the model (iii) the direct detections 
would be feasible for $\xi>-0.013$ (upgraded DECIGO) and $\xi>-0.043$ (BBO).

We have thus shown that the large-field inflationary models 
with $r$ larger than the order of 0.01 leave interesting 
observational signatures in the gravitational wave background. 
In particular, the models in which reheating is driven by an 
effective massless inflaton field can be distinguished from 
the massive models with a temporal matter era after inflation. 
We hope that, in addition to the indirect detection of tensor 
perturbations from the B-mode polarization, 
the measurements by interferometric detectors 
will shed new light on the nature of primordial gravitational waves 
in the foreseeable future.

\section*{ACKNOWLEDGEMENTS}

This work is supported by the Grant-in-Aid for Scientific Research
from JSPS (Nos.\,24540287 (TC), 24540286 (ST), 25287057(NS)), by the
Grant-in-Aid for Scientific Research on Innovative Areas (No.~21111006
(ST)) and in part by Nihon University (TC) and by World Premier
International Research Center Initiative (WPI Initiative), MEXT,
Japan. Compared to the version published in Physical Review D, 
we also incorporated the sensitivity curves and the signal-to-noise ratios 
of Ultimate DECIGO in this updated arXiv version.
We thank Takashi Nakamura for his suggestion to include the 
discussion of Ultimate DECIGO.


\end{document}